%% file: 00.tex
\documentclass[12pt,a4paper]{article}

\newcommand{\citeN}[1]{\cite{#1}}
\usepackage[ruled,vlined,linesnumbered]{algorithm2e}
\usepackage{algorithmicx}
\usepackage{mathtools,amssymb,amsthm,pgf,tikz,xspace,units,color,enumitem,footnote,changes,anysize}
\usepackage{todonotes}

\renewenvironment{proof}{\begin{oldproof}}{\qed\end{oldproof}}

\marginsize{1in}{1in}{1in}{1in}

\newcommand{\RR}{\ensuremath{\mathbb{R}}}
\newcommand{\bigO}{\mathcal{O}}

\renewcommand{\Pr}[1]{\mbox{\rm\bf Pr}\left[#1\right]}
\newcommand{\Ex}[1]{\mbox{\rm\bf E}\left[#1\right]}

\newcommand{\OPT}{\mathrm{OPT}}

\newcommand{\classNP}{{\sf NP}}
\newcommand{\classFPT}{{\sf FPT}}
\newcommand{\classWOne}{{\sf W[1]}}

\newcommand{\xiS}{x_{i,S}}
\newcommand{\qiS}{q_{i,S}}
\newcommand{\qiE}{q_{i,\emptyset}}
\newcommand{\vi}{w_i}
\newcommand{\vj}{w_j}
\newcommand{\ep}{\epsilon}
\newcommand{\Ks}{K^*}
\newcommand{\polymn}{\textsf{poly}$(m,n)$}

\newcommand{\swu}{{SW}_u}
\newcommand{\sw}{{SW}}
\newcommand{\swe}{{SW}}
\newcommand{\rd}{R(\Delta)}

\newcommand{\vup}{\vspace*{-0.03in}}

\newcommand{\orderone}{\phi}
\newcommand{\ordertwo}{\chi}

\newtheorem{theorem}{Theorem}
\newtheorem{lemma}[theorem]{Lemma}
\newtheorem{proposition}[theorem]{Proposition}
\newtheorem{corollary}[theorem]{Corollary}

\newtheorem{claim}[theorem]{Claim}

\newtheorem{definition}{Definition}

\newenvironment{pf}{\begin{proof}[\emph{\textbf{Proof: }}]}{\end{proof}}
\newenvironment{pfof}[1]{\begin{proof}[\emph{\textbf{Proof of #1: }}]}{\end{proof}}

\title{Combinatorial Auctions with Conflict-Based Externalities\thanks{This work
was funded by the Vienna Science and Technology Fund (WWTF) through project ICT10-002.
This work is supported by DFG through Cluster of Excellence MMCI.}}
\author{
Yun Kuen Cheung\thanks{University of Vienna, Faculty of Computer Science.}
\and
Monika Henzinger\footnotemark[2]
\and
Martin Hoefer\thanks{Max-Planck-Institut f\"ur Informatik and Saarland University.}
\and
Martin Starnberger\footnotemark[2]
}
\date{}

\begin{document}

\maketitle

\normalem

\input{abstract}

\input{introduction}
\input{relatedwork}
\input{Rounding-var}
\input{cone_program}
\input{SubLinSponsoredSearch}

\input{sponsored_search_limited_number_slots}


\bibliographystyle{alpha}
\bibliography{literature}

\newpage
\appendix

\input{Rounding-Appendix}

\input{cone_program_appendix}

\end{document}

%% file: abstract.tex
\begin{abstract}
Combinatorial auctions (CA) are a well-studied area in algorithmic mechanism design.
However, contrary to the standard model, empirical studies suggest that a bidder's valuation often
does not depend solely on the goods assigned to him.
For instance, in adwords auctions an advertiser might not want his ads to be displayed next to his competitors' ads.
In this paper, we propose and analyze several natural graph-theoretic models that incorporate such negative externalities, in which bidders form a directed conflict graph with maximum out-degree $\Delta$.
We design algorithms and truthful mechanisms for social welfare maximization
that attain approximation ratios depending on $\Delta$.

For CA, our results are twofold:
(1)~A \emph{lottery} that eliminates conflicts by discarding bidders/items independent of the bids.
It allows to apply any truthful $\alpha$-approximation mechanism for conflict-free valuations and yields an $\bigO(\alpha\Delta)$-approximation mechanism.
(2)~For fractionally sub-additive valuations, we design a rounding algorithm via
a novel combination of a semi-definite program and a linear program, resulting in a \emph{cone program};
the approximation ratio is $\bigO((\Delta \log \log \Delta)/\log \Delta)$.
The ratios are almost optimal given existing hardness results.

For the prominent application of adwords auctions,
we present several algorithms for the most relevant scenario when the number of items is small.
In particular, we design a truthful mechanism with approximation ratio $o(\Delta)$ when the number of items is only logarithmic in the number of bidders.
\end{abstract}

%% file: introduction.tex
%
\section{Introduction}

Combinatorial auctions (CA) are an important area in algorithmic mechanism design
due to wide-spread applications in resource allocation and e-commerce, e.g., spectrum or adwords auctions~\cite{CramtonSS06}.
In the standard CA, a set of items is assigned to a set of bidders in order to maximize social welfare,
which is given by the total valuations of bidders for their assigned items.
This assumes that each bidder values exclusively the set of items assigned to him ---
his valuation is \emph{independent of the assignment of other items to other bidders}.
In many applications (see \cite{JehielMS96,JehielMS99} for examples), however,
such an assumption is not justified since bidder preferences have a significant dependence
on how items are assigned to other bidders. Such a dependence is called \emph{externality}.

Mechanism design for CA with externalities in the most general form is difficult,
primarily due to the huge complexity of bidders' preferences with externalities,
which then also leads to the computational complexity issue for (approximately) maximizing social welfare.
Prior work has studied more restricted scenarios, e.g., when there is only one item on sale,
or when the bidders' preferences are simple (e.g., unit-demand).
In this work, we focus on a simple type of externalities called \emph{conflict-based externality},
which is readily motivated by sponsored search auctions (SSA);
in our model, there are multiple items on sale, and the bidders' preferences might be more complex than unit-demand ones.

SSA are one of the most popular special cases of CA, where ad slots on a search result page are assigned to advertisers.
\emph{Negative externality} arises when, for example, a car-rental company has much smaller value for an ad slot
if an ad of another prominent rental company is shown right next to it.
More generally, for an advertiser there might be a number of competitors,
and an assignment yields value to the bidder only if the ads of competitors are not displayed simultaneously.
The existence of negative externalities in sponsored search has been confirmed empirically~\cite{GomesIM09}.
Moreover, similar negative externalities also arise in other prominent applications of CA,
e.g., in secondary spectrum auctions where interferences induce negative externalities;
or when selling luxury goods, where the value of a buyer for items from an exclusive brand drops
when other buyers also obtain items from the same brand.
These examples give rise to a natural and simple graph-based model of externalities:
each bidder is a node in a directed graph, and a directed edge indicates that a bidder sees another bidder as a competitor;
assigning an item to a bidder yields value only if none of the competitors receives any item (or just any ``similar'' or ``better'' item).

Negative externalities in auctions have recently received attention, but --- perhaps surprisingly ---
the natural and simple idea sketched above has not been analyzed in a rigorous and general fashion.
We propose three graph-theoretic models that incorporate these conflict-based externalities.
We study approximation algorithms and truthful mechanisms under the models.
Formally, we assume there is a directed conflict graph on the set of bidders.
Each edge $(i,j)$ indicates a conflict: $i$ has no value for any assignment in which $j$ receives an item.
More generally, we also consider cases where conflicts arise only among certain pairs of items,
or different values for assignments that include or avoid certain conflicts.
Our algorithms cope with externalities via new extensions of algorithmic techniques for independent set problems
{\em in combination with} algorithms for conflict-free CA.
We also provide additional results for the prominent special case of SSA.
Before we state our results, we proceed with a formal introduction and discussion of
the models on conflict-based externalities treated in this paper.

\subsection{Auctions with Conflict-Based Externalities}

In all models, we have a bidder set $B$ of $n$ bidders and an item set $I$ of $m$ items.
Each item can be given to at most one bidder.
For each $i \in B$, there is a valuation function $v_i:\,2^I\rightarrow \RR^+$,
where $v_i(S_i)$ represents the value for receiving item set $S_i \subseteq I$.
In the SSA case, the items in $I$ are ad slots. Each slot $k$ has a click-through rate $\alpha_k\geq 0$.
Each bidder $i$ has a valuation per click of $v_i\geq 0$ in one slot.
Then $v_i(S_i) = \max_{k\in S_i} v_i \cdot \alpha_k$, a unit demand valuation function with free disposal.

The valuation $v_i(S_i)$ will be extended, due to externalities, to $v_i^c(S)$,
a valuation that depends on the complete allocation $S=(S_1,\cdots,S_n)$.
The goal is to find an allocation $S$ that maximizes social welfare $SW(S) = \sum_{i\in B} v_i^c(S)$.

\medskip

\noindent\underline{CA with Bidder Conflicts.}
The set of bidders $B$ is the vertex set of a {\em (bidder) conflict graph} $G = (B,E)$, which is a directed graph.
Each bidder $i$ has a valuation function $v_i:\,2^I\rightarrow \RR^+$.
Given a complete allocation $S=(S_1,\cdots,S_n)$,
\begin{eqnarray*}
	v_i^c(S) = \begin{cases}
	v_i(S_i) & \text{ if } \bigcup_{j : (i,j) \in E} S_j = \emptyset\\
	0 & \text{ otherwise. }
	\end{cases}
\end{eqnarray*}
This models the situation that advertiser $i$ is not interested in showing its ad
together with an ad from a competitor $j$, represented by an edge $(i,j)\in E$.

The introduction of conflicts turns social welfare maximization \classNP-hard; in
the special case SSA with all $v_i=1$, all $\alpha_k=1$, and $m=n$,
it reduces to the maximum independent set problem.

\medskip

\noindent\underline{CA with Bidder and Item Conflicts.}
There are two conflict structures in this model, each represented by a directed graph.
The bidder set $B$ is the vertex set of a \emph{bidder conflict graph} $G = (B,E)$.
The item set $I$ is the vertex set of an \emph{item conflict graph} $G_I = (I,E_I)$.
Both graphs are directed.
Intuitively speaking, if $(i,j)\in E$ and $(k,\ell)\in E_I$,
then bidder $i$ has no use for item $k$ if $j$ receives item $\ell$.
Formally, for any allocation $S$, bidder $i$ has a set $D_i$ of \emph{useless items}, defined as
$D_i := \{ k \in S_i \mid \exists \ell \in S_j : (i,j) \in E \text{ and } (k,\ell) \in E_I\}$,
and $v_i^c(S) := v_i(S_i \setminus D_i)$.

An intuitive example is \emph{ordered conflicts},
where ad slots are ordered on a page top-down,
and a bidder has a conflict only if a competitor receives a slot \emph{above} him.
This can be modelled by numbering slots top-down and
$E_I = \{ (k,\ell) \mid k,\ell \in I, \ell < k \}$.
Another intuitive example is \emph{neighbor conflicts},
where ad slots are arranged horizontally,
and a bidder has conflict only if a competitor receives a slot \emph{right next to} him.
This can be modelled by numbering slots from left to right and
$E_I = \{ (k,\ell) \mid k,\ell \in I, |k - \ell| = 1 \}$.

Note that CA with bidder conflicts is a sub-case of this model, when $G_I$ is the complete digraph.

The results in this paper depend on two parameters $\Delta$ and $\Delta_I$ of the conflict graphs,
which are the maximum out-degrees of the graphs $G$ and $G_I$ respectively.

\medskip

\noindent\underline{CA with Bidder Conflicts and Conflict Value.}
In CA with bidder conflicts, we assume that the valuation of a bidder drops to $v_i(S) = 0$
as soon as a competitor receives any item.
We can generalize this assumption to a second valuation function $w_i(S_i)$:
if $\bigcup_{j : (i,j) \in E} S_j = \emptyset$, then $v_i^c(S) = v_i(S_i)$;
otherwise, $v_i^c(S) = w_i(S_i)$.

This model can be reduced to the model with bidder-conflicts only.
Given an instance of CA with bidder conflicts and conflict value,
we build an instance without conflict value as follows:
for each bidder $i$, we add an auxiliary bidder $i_c$, where $v_{i_c}(S_{i}) = w_i(S_i)$.
In the bidder conflict graph, we add the edges $(i,i_c)$ and $(i_c,i)$. This increases $\Delta$ by exactly 1.
Now if bidder $i$ is conflicted, we can take all items assigned to it and assign them to bidder $i_c$ instead.
In this way, we can transform any allocation into the instance without conflict value and obtain the same social welfare.
It is straightforward to observe that social welfare maximization in both instances is equivalent.
This, however, does not directly apply to truthfulness.

\smallskip

There are numerous further ways to extend our models, e.g., to combinations of item conflicts and conflict values,
weighted conflicts, etc. Studying their properties are interesting avenues for future work.

\subsection{Our Contribution}

For CA with conflict-based externalities, we design and analyze poly-time approximation algorithms and truthful mechanisms
which provide almost best possible approximation guarantees of maximizing social welfare.
To state our results, we first define the class of \emph{fractionally sub-additive} valuations,
which is known to strictly contain the more well-known unit-demand valuations, linear valuations,
gross substitute valuations and submodular valuations.

\begin{definition}[See \cite{FeigeV10}]
A fractionally sub-additive valuation is a valuation function $v:2^I\rightarrow \mathbb{R}$ that satisfies the following property
for any $S,T_1,T_2,\cdots,T_k \in 2^I$ and $0\le \alpha_1,\alpha_2,\cdots,\alpha_k \le 1$:
if for all $j\in S$, $\sum_{\ell:~j\in T_\ell} \alpha_\ell \geq 1$, then
$v(S) \leq \sum_{\ell=1}^k \alpha_\ell \cdot v(T_\ell)$.
\end{definition}

For CA with bidder conflicts, we use well-known techniques for independent set problem
to eliminate conflicts, which is in a spirit similar to \emph{lottery}, to give a reduction to conflict-free CA.
Given any $\alpha$-approximation algorithm for the unconflicted problem,
we obtain an $\bigO(\alpha \Delta)$-approximation algorithm for CA with bidder conflicts (Theorem \ref{thm:randmain1}).
If the original algorithm is a truthful mechanism, our reduction preserves the truthfulness.
Moreover, our reduction preserves the use of randomization (deterministic, universally truthful, truthful in expectation).
If the bidders have fractionally sub-additive valuations,
our results extend to CA with bidder and item conflicts (Theorem \ref{thm:randmain2}).

The next natural question to ask is whether one can improve the approximation ratio to $o(\Delta)$.
Since our problem generalizes the weighted independent set (WIS) problem\footnote{Given any graph $G=(V,E)$,
$J\subset V$ is an independent set if no two vertices in $J$ are connected by an edge in $E$.
The WIS problem is: suppose each vertex $v\in V$ has a positive weight $w_v$,
find an independent set $J$ which maximizes $\sum_{v\in J} w_v$.},
the ratio must be $\Omega(\Delta/\log^4 \Delta)$~\cite{Chan13}, even for unit-demand valuations.
We answer the question positively:
if the bidders have fractionally sub-additive valuations and if there is a \emph{demand oracle} for each bidder,
we design an $\bigO((\Delta \log \log \Delta)/\log \Delta)$-approximation algorithm (Theorem \ref{thm:CP-main}).
This implies, for example, ratios of $\bigO((\Delta \log \log \Delta)/\log \Delta)$ for sponsored search,
unit-demand, or more general gross-substitute valuations.
The dependence on $\Delta$ mirrors the best-known approximation ratio for WIS.
Our algorithm combines an approach for WIS based on semi-definite programming (SDP)
with the standard approach for CA based on linear programming (LP) to design a cone program relaxation and a rounding scheme.
To the best of our knowledge, we are the first to combine an SDP with an LP in this fashion, and to show how to analyze it.
We believe this technique might be of independent interest in other applications.
It is an interesting open problem if this approach can be turned into a truthful mechanism,
or be generalized to CA with bidder and item conflicts.

We then focus on SSA with bidder conflicts.
Even in this special case, the hardness bound of $\Omega(\Delta/\log^4 \Delta)$ applies.
We consider a restriction to a small number of slots that is natural in the context of sponsored search.
For the case of $m = \bigO(\log n)$ slots, we present a truthful mechanism based on SDP
that obtains an $\bigO(\Delta \cdot \sqrt{(\log \log \Delta)/(\log \Delta)})$-approximation (Theorem \ref{thm:ssmainthm}).
To obtain the desired truthfulness property,
the first step of our mechanism is to gather a statistic from a sampling of bidders who will not be allocated any item,
which is similar the first two steps in the framework of Dobzinski et al.~\cite{DobzinskiNS12} for designing truthful mechanisms.
However, the subsequent steps of our algorithm will be different from theirs.

Also, we get an $\bigO(\log m)$-approximation algorithm based on partial enumeration
that runs in time $\bigO((m\Delta)^m)$ (Theorem \ref{thm:small-supply});
the algorithm can be turned into truthful-in-expectation mechanisms with the same approximation guarantee,
and it extends to CA with bidder and item conflicts.

%% file: relatedwork.tex
\subsection{Related Work}

The study of auctions with externalities was initiated by seminal work of Jehiel et al.~\citeN{JehielMS96,JehielMS99} in the single-item setting.
The externality in this work is \emph{identity-dependent}, i.e., each bidder can have a different valuation when different bidders obtain the item.
The preference of each bidder can thus be represented by a low dimensional $\mathbb{R}^{n+1}$ vector,
which reflects the bidder's valuation on the $(n+1)$ possible outcomes.
In our model, a bidder is indifferent between the bidders who he conflicts with, but our model allows multiple items in an auction.

Gomes et al.~\cite{GomesIM09} gave empirical evidence that externalities exist in real-life SSA.
Externalities in online advertising were investigated by~\cite{GhoshM08} using a probabilistic model.
CA with externalities were presented in~\cite{KrystaMSW10,ConitzerS12,HaghpanahIMM13}, and maximizing social welfare was shown to be \classNP-hard.
In~\citeN{GhoshS10} a sponsored search setting was treated where each advertiser has two valuations,
one if his ad is shown exclusively and one if it is shown together with other ads. 
This is a special case of our model for CA with bidder conflicts and conflict values.
A different line of work considered bidder-independent externalities in the click-through rates of SSA~\cite{AggarwalFMP08,KempeM08,RoughgardenT12}.
All this work considered only the unit-demand setting.

Our model of SSA with bidder conflicts has been proposed and studied before by Papadimitriou and Garcia-Molina~\cite{PapadimitriouG12}.
They consider an approach based on exact optimization algorithms using ILP, implement truthfulness using VCG,
and experimentally evaluate their approach with respect to running time and revenue on a dataset from the Yahoo!\ Webscope.
However, they do not consider polynomial-time algorithms, provable approximation ratios, or extensions to CA with more general valuations.

Our work is related to approximation algorithms for weighted independent set problem,
a central problem in the study of approximation algorithms and computational hardness over the past four decades.
For a survey on some of the work on approximation algorithms, see, e.g.,~\cite{Halldorsson98};
here, we just mention a number of directly related results.
The problem is known to be \classNP-hard to approximate within a ratio of $n^{1-\epsilon}$~\cite{Hastad99},
and even in undirected $\Delta$-regular graphs it remains hard for a ratio of $\bigO(\Delta/\log^4 \Delta)$~\cite{Chan13}.
A trivial greedy algorithm obtains an approximation ratio of $(\Delta+1)$ in undirected graph with maximum degree $\Delta$.
For directed graphs, which arise in our application, a simple randomized $(4\Delta)$-approximation algorithm exists.
The best-known approximation algorithms for undirected graphs with maximum degree $\Delta$ attain ratios of
$\bigO((\Delta \log \log \Delta)/\log \Delta)$~\cite{Halldorsson00,Halperin02}. They are based on rounding suitable SDP relaxations,
and below we build on these techniques and their analysis to provide algorithms for our cases, which involve directed graphs.

More recently, the study of asymmetric and edge-weighted versions of independent set has found interest,
especially in the context of secondary spectrum auctions~\cite{ZhouGSZ08,HoeferK12,HoeferK13,HoeferKV14},
where bidders are wireless devices that strive to obtain channel access under interference constraints.
In these scenarios, bidders become vertices in a conflict graph. Each channel is an item that can be given to any subset of bidders
representing an independent set in the graph.

%% file: Rounding-var.tex

\section{CA with Bidder and Item Conflicts via Lottery}\label{sec:randmech}\label{SEC:RANDMECH}

In this section, we present results for CA with bidder and item conflicts.
We assume that either (i) $\Delta$ is bounded and $G_I$ is arbitrary,
or (ii) $\Delta_I$ is bounded, $G$ is arbitrary and bidders have fractionally sub-additive valuations.

\subsection{Bounded Out-degree in the Bidder Conflict Graph}

For case (i), we prove the following result:

\begin{theorem}
\label{thm:randmain1}
Given a (maximal-in-range) deterministic $\alpha$-approximation algorithm $f$ for CA without conflicts,
there exists a (truthful maximal-in-range) deterministic $(16\Delta\alpha/3)$-approximation algorithm $f^c$
for CA with bidder and item conflicts satisfying condition (i).
\end{theorem}

The main idea of Theorem~\ref{thm:randmain1} is to first generate a ``good'' conflict-free bidder set $B^c$,
and then apply the blackbox algorithm $f$ w.r.t.~the bidders in $B^c$.
Initially, each bidder is in $B^c$ with probability $1/(2\Delta)$.
Then, if there are still bidder conflicts within $B^c$, we remove those bidders having conflicts.
Overall, each bidder is in $B^c$ with probability of at least $1/(4\Delta)$,
and this will translate into a randomized $(4\Delta\alpha)$-approximation algorithm $f^c$.

We derandomize the above algorithm using the standard technique of pairwise independent distributions~\cite{LubyW05};
this will lead to an increase of the approximation ratio from $4\Delta\alpha$ to $16\Delta\alpha/3$.
Since no bidder can alter $B^c$ by changing his valuation, truthfulness is preserved from $f$ to $f^c$.
The details of derandomization is given in Appendix \ref{subsect:sampling-derandomization}.

Next, we discuss the details of designing the randomized $(4\Delta\alpha)$-approximation algorithm.
The results apply to arbitrary restrictions on the valuations (e.g., submodular valuations).
Given an allocation $S=(S_1,\dots,S_n)$ of items to bidders,
we show how to compute a random set $B^c\subseteq B$ such that
(1)~if $(i,j)\in E$ then $i\not\in B^c$ or $j\not\in B^c$,
(2)~$\sum_{i\in B} v_i(S)\le (4\Delta)\mathbf{E}_{B^c}[\sum_{i\in B^c} v_i(S)]$, and
(3)~the selection of $B^c$ does not depend on the valuations.

We will use pairwise independent distributions;
such distributions always exist as one can pick the elements in $B$ independently with probability $q$.
\begin{definition}
We call a distribution $\mathcal{D}$ over subsets of a set $B$ ``pairwise independent with probability $q$'' if for $B^R\sim \mathcal{D}$ and $i\neq j\in B$ holds that $\Pr{i\in B^R}=q$ and $\Pr{\{i,j\}\subseteq B^R}=\Pr{i\in B^R}\cdot\Pr{j\in B^R}$.
\end{definition}

\begin{algorithm}[h]
\SetAlgoLined
Pick a random subset $B^R$ from a distribution over subsets of $B$ that is pairwise independent with probability $1/(2\Delta)$\;\label{ln:step1}
$B^c\gets B^R$\;
\textbf{For each} $i\in B^R$:\ \textbf{if} $\exists j\in B^R$ with $(i,j)\in E$ \textbf{then} delete $i$ from $B^c$\;\label{ln:step2}
\KwRet{$B^c$}
\caption{Conflict-free random set}
\label{alg:randset}
\end{algorithm}

The random set $B^c$ computed by Algorithm~\ref{alg:randset} is constructed in the following way.
First, in line~\ref{ln:step1} the algorithm picks a random subset from a pairwise independent distribution with probability $1/(2\Delta)$.
Next, in line~\ref{ln:step2} the algorithm resolves all the remaining conflicts between the bidders.
The proof of Lemma~\ref{thm:detalg} exploits that every bidder is in $B^c$ with probability at least $1/(4\Delta)$.

\begin{lemma}
\label{lem:randset}
In Algorithm~\ref{alg:randset}, every bidder is in $B^c$ with a probability of at least $1/(4\Delta)$.
\end{lemma}
\begin{pf}
For all $i\in B$ let $Q_i$ be the event that $i\in B^R$ in line~\ref{ln:step1}.
Thus, the probability for this event is $\Pr{Q_i}=1/(2\Delta)$.
The probability that a bidder $i\in B^R$ gets deleted in line~\ref{ln:step2}, conditioned on that it was selected in $B^R$,
is $\Pr{\bigcup_{j\in N(i)}Q_j\mid Q_i}$, where $N(i)$ denotes the set of out-neighbors of bidder $i\in B$.
\begin{multline*}
\Pr{\bigcup_{j\in N(i)}Q_j\Biggm{|}Q_i}
=\frac{\Pr{(\bigcup_{j\in N(i)}Q_j)\cap Q_i}}{\Pr{Q_i}}
=\frac{\Pr{\bigcup_{j\in N(i)}(Q_j\cap Q_i)}}{\Pr{Q_i}}\\
\overset{(*)}{\le} \frac{\sum_{j\in N(i)}\Pr{Q_j\cap Q_i}}{\Pr{Q_i}}
\overset{(**)}{=} \frac{\sum_{j\in N(i)}\Pr{Q_j}\Pr{Q_i}}{\Pr{Q_i}}
= \sum_{j\in N(i)}\Pr{Q_j}
\le \frac{1}{2}.\,
\end{multline*}
In the above (in)equalities, $(*)$ follows from Boole's inequality and $(**)$ follows from pairwise independence.

Thus, the probability that $i\in B^c$ at the end of the loop is
\begin{equation*}
\Pr{Q_i}\Pr{\bigg(\bigcup_{j\in N(i)}Q_j\bigg)^\mathrm{C}\Biggm{|}Q_i}=\Pr{Q_i}\left(1-\Pr{\bigcup_{j\in N(i)}Q_j\Biggm{|}Q_i}\right)\ge \frac{1}{4\Delta}\,.
\end{equation*}
\end{pf}

\begin{lemma}
\label{thm:detalg}
A (randomized) $\alpha$-approximation algorithm $f$ for CA without conflicts
can be turned into a randomized 
$(4\Delta\alpha)$-approximation algorithm $f^c$ for CA with bidder and item conflicts, in which $\Delta$ bounded and $G_I$ arbitrary.
\end{lemma}
\begin{pf}
We define our $(4\Delta\alpha)$-approximation algorithm $f^c$ as follows:
$f^c$ first calls Algorithm~\ref{alg:randset} to compute a random subset $B^c$
and then calls $f$ for the bidders in $B^c$.

Assume that the $\alpha$-approximation algorithm $f$ returns the allocation\linebreak $(S_1(B'),\dots,S_n(B'))$
if we use it on the set of bidders $B'\subseteq B$ and if we set $S_i(B')=\emptyset$ for all $i\not\in B'$.
Furthermore, assume that the optimal allocation is $(\OPT_1(B'),\dots,\OPT_n(B'))$ given the constraints $\OPT_i(B')=\emptyset$ if $i\not\in B'$.
Then
\begin{multline*}
\mathbf{E}_{B^c}\bigg[\sum_{i\in B^c}v_i(S_i(B^c))\bigg]\ge \mathbf{E}_{B^c}\bigg[\frac{1}{\alpha}\sum_{i\in B^c} v_i(\OPT_i(B^c))\bigg]\\ \overset{(*)}{\ge}\frac{1}{\alpha}\mathbf{E}_{B^c}\bigg[\sum_{i\in B^c} v_i(\OPT_i(B))\bigg]=\frac{1}{\alpha}\sum_{i\in B} v_i(\OPT_i(B))\cdot\Pr{i\in B^c}\\ \ge\frac{1}{4\Delta\alpha}\sum_{i\in B}v_i(\OPT_i(B)).
\end{multline*}
$(*)$ holds because $\OPT(B^c)$ gives the bidders in $B^c$ the maximal social welfare.
\end{pf}

Since no bidder can alter $B^c$ by changing his valuation,
it also holds that a universally truthful (resp.~truthful in expectation) $\alpha$-approximation mechanism $(f,p)$
for combinatorial auctions without conflicts
can be turned by the same approach into a universally truthful (resp.~truthful in expectation)
$(4\Delta\alpha)$-approximation mechanism $(f^c,p^c)$ for combinatorial auctions with conflicts.
That is, we can first call Algorithm~\ref{alg:randset} to compute a random subset $B^c$
and can then use $(f,p)$ only for the bidders in $B^c$.
Note that bidders not in $B^c$ cannot change their utility by changing their bid; it is always zero.
Furthermore, bidders in $B^c$ behave like in $(f,p)$, i.e.~they bid truthfully so as to maximize their own utilities
for each realization of $B^c$.
The approximation guarantee follows from Lemma~\ref{thm:detalg}.
\begin{corollary}\label{coro:truthful}
A universally truthful (resp.~truthful in expectation) $\alpha$-approximation mechanism for CA without conflicts
can be turned into a universally truthful (resp.~truthful in expectation)
$(4\Delta\alpha)$-approximation mechanism for CA with $\Delta$ bounded and $G_I$ arbitrary.
\label{cor:randmech}
\end{corollary}

\subsection{Bounded Out-degree in the Item Conflict Graph}

For case (ii), the ideas are similar, but here we generate a conflict-free item set $I^c$ using the same technique as in case (i).
We require that bidders have fractionally sub-additive valuations to avoid the existence of complementary goods,
which should not be deleted independently as the following example illuminates:
Suppose there are two items which are \emph{total complements} and they have item conflicts.
If we remove either of the two, the social welfare drops to zero,
while the optimal allocation attains a positive social welfare by allocating both items to the same bidder.

The algorithm for case (ii) and its analysis are similar to those for case (i),
except that we need the following lemma about fractionally sub-additive valuations,
which will imply that if each item is not discarded with probability at least $1/k$,
then social welfare retained is at least $1/k$ of the original.
For comparison, for case (i), we have shown that if each bidder is not discarded with probability at least $1/k$,
then the eventual social welfare retained is at least $1/k$ of the original.

\begin{lemma}[{\cite[Proposition 2.3]{Feige09}}]
\label{prop:feige}
Let $k \ge 1$ and let $w$ be an arbitrary fractionally sub-additive utility function.
For a set $S$, consider a distribution over subsets $S' \subset S$ such that
each item of $S$ is included in $S'$ with probability at least $1/k$. Then $\mathbf{E}[w(S')] \ge w(S)/k$.
\end{lemma}

We obtain the following theorem; we defer its proof to Appendix \ref{subsect:sampling-item}.

\begin{theorem}\label{thm:randmain2}
For CA with bidder and item conflicts satisfying conditions (i) and (ii),
given a (maximal-in-range) deterministic $\alpha$-approximation algorithm $f$ for CA without conflicts,
there exists a (truthful maximal-in-range) deterministic
$(16\alpha/3)\cdot \min\{\Delta,\Delta_I\}$-approximation algorithm $f^c$.
\end{theorem}

%% file: cone_program.tex
\section{CA with Bidder Conflicts via Cone Program Relaxation}\label{SECT:CP}\label{sect:CP}

\newcommand{\LPR}{\mbox{\textsf{LPR}}}
\newcommand{\FCRA}{\mbox{\textsf{FCRA}}}
\newcommand{\hatx}{\hat{x}}

We design an approximation algorithm via a combination of
(i) an SDP for WIS problem and (ii) an LP for conflict-free CA.
This yields a \emph{cone program}, which we round its solution to yield a good allocation.
Cone program (CP) is a generalization of the more well-known LP and SDP.
Briefly speaking, a CP is a program that optimizes a linear function
in the intersection of some hyperspaces and a \emph{proper cone}. More about CP will be given in Section \ref{subsect:CP-cp}.

To the best of our knowledge, we are the first to combine this SDP and a LP in this fashion.
Also, our analysis for the cone program rounding algorithm is novel,
combining of the analyses for (i) and (ii) so as to get the benefits of both.
We prove the following result:

\begin{theorem}\label{thm:CP-main}
For CA with bidder conflicts, suppose that the bidders have fractionally sub-additive (FSA) valuations.
If there is a demand oracle for each bidder (we shall define this in Section \ref{subsect:CP-CA-noconflict}),
then there exists an $\bigO\left((\Delta \log\log \Delta)/\log \Delta\right)$-approximation algorithm
of social welfare that runs in \emph{\polymn}-time.
\end{theorem}

The rest of this section is organized as follows.
We include the relevant standard facts about CA without conflicts in Section \ref{subsect:CP-CA-noconflict}.
We include the formal definition of cone program and the relevant standard facts about cone programs in Section \ref{subsect:CP-cp}.
Then we describe our algorithm in Section \ref{subsect:CP-algorithm}, 
provide some intuitions of the algorithm in Section \ref{subsect:CP-intuition},
and give its analysis in Section \ref{subsect:CP-analysis}.
We have a few remarks and discussion in Section \ref{subsect:CP-discussion}.
We slightly abuse the notation and use $S$ to denote a subset of $I$.


\subsection{CA with No Conflicts}\label{subsect:CP-CA-noconflict}

The optimal social welfare of a CA without conflicts can be represented
by the program \textsf{ILP-NC}: maximizing $\sum_{i\in B} \sum_{S\neq\emptyset} v_i(S) \cdot \xiS$
subject to three sets of constraints:
(i) $\forall i\in B,~\sum_{S\neq\emptyset} \xiS \leq 1$;
(ii) $\forall k\in I,~\sum_{S\ni k} \sum_{i\in B} \xiS \leq 1$;
(iii) $\forall i\in B$ and $\forall S\subseteq I$, $\xiS\in\{0,1\}$.


In general, solving \textsf{ILP-NC} is \classNP-hard.
The usual remedy is to solve its LP relaxation \textsf{LPR-NC},
i.e., relaxing (iii) from $\xiS\in\{0,1\}$ to $\xiS\in [0,1]$, to obtain a \emph{fractional} solution,
and \emph{round} it to an integral solution.
%

There are $\Omega\left(2^m n\right)$ variables in \textsf{LPR-NC}, but it can be solved in \polymn-time
if there is a \emph{demand oracle} for each bidder: given the prices of the items $p_1,p_2,\cdots,p_m$,
the demand oracle of bidder $i$ returns a set $S\subseteq I$ that maximizes $v_i(S) - \sum_{k\in S} p_k$.
The demand oracles serve as \emph{separation oracles} for the dual of \textsf{LPR-NC},
thus allow solving \textsf{LPR-NC} efficiently using the ellipsoid algorithm~\cite{NisanS06}.

For CA without conflicts where bidders have FSA valuations, a rounding algorithm called
\emph{fair contention resolution algorithm} (FCRA)~\cite[Section 1.2]{FeigeV10} attains approximation ratio $1-\frac{1}{e}$.
In Lemma \ref{lem:FCRA} below, we state the precise result on FCRA, which will be useful for our conflict setting;
we need the following notation:
$\forall B'\subseteq B$, let $\LPR (B')$ denote the program \textsf{LPR-NC} with the item set $I$ and the bidder set restricted to $B'$.
Given any feasible point $\{\xiS\}_{i\in B',S\subseteq I}$ of $\LPR (B')$, $\forall i\in B'$,
let $L_i\left(\{\xiS\}_{i\in B',S\subseteq I}\right) := \sum_{S\neq\emptyset} v_i(S)\cdot \xiS$.

\begin{lemma}[\cite{FeigeV10}]\label{lem:FCRA}
Suppose the bidders in $B$ have FSA valuations. Given any feasible point $\{\xiS\}_{i\in B,S\subseteq I}$ of $\LPR (B)$,
FCRA outputs a randomized allocation in which each $i\in B$ obtains expected welfare of at least
$\left(1-\frac{1}{e}\right)\cdot L_i\left(\{\xiS\}_{i\in B,S\subseteq I}\right)$.
\end{lemma}

Let $\FCRA \left(B,\{\xiS\}_{i\in B,S\subseteq I}\right)$ denote the randomized allocation in Lemma \ref{lem:FCRA}.
For any $B'\subseteq B$, let $\hatx(B')$ denote the optimal solution to $\LPR(B')$.


\subsection{Cone Programs in a Nutshell}\label{subsect:CP-cp}

Cone programs (CP) are generalization of the more familiar linear programs (LP) and semi-definite programs (SDP). We list the relevant definitions and properties of CP here. They are extracted from Chapters 2 and 5 in \cite{BoydV04}.

A closed set $K \subseteq \mathbb{R}^q$ is a \emph{proper cone} if
(a) for any real numbers $a_1,a_2\geq 0$ and for any $k_1,k_2\in K$, $a_1 k_1 + a_2 k_2 \in K$;
(b) it has nonempty interior; and
(c) it is pointed, i.e., if $x,-x\in K$, then $x=0$.
Two examples of proper cones are the non-negative orthant (the set of points with non-negative coordinates),
and the set of symmetric positive semi-definite (SPSD) matrices.

There is a natural partial ordering on $\mathbb{R}^q$ associated with any proper cone $K$, which is denoted by $\succeq_K$:
For any $x_1,x_2\in \mathbb{R}^q$, $x_1 \succeq_K x_2$ if and only if $(x_1 - x_2) \in K$.
The corresponding \emph{strict} partial ordering, $\succ_K$, is defined as follows:
For any $x_1,x_2\in \mathbb{R}^q$, $x_1 \succ_K x_2$ if and only if $(x_1 - x_2)$ is an interior point of $K$.

The \emph{dual cone} of a proper cone $K$, is the set $\Ks := \left\{z~\,|\,~\forall k\in K,~k \cdot z\geq 0\right\}$, where $k\cdot z$ is the inner product of $k$ and $z$. The dual cone of the non-negative orthant is itself, and the dual cone of the set of SPSD matrices is again itself~\cite[Examples 2.23 and 2.24]{BoydV04}. The two proper cones are said to be \emph{self-dual}.

There are various forms of CP, but they can be shown to be equivalent. In this paper, we will use two of the forms of CP. The \emph{standard form} of CP is as follows; note that $c,x\in \mathbb{R}^q$, $A$ is an $\ell \times q$ real matrix, $b\in \mathbb{R}^\ell$ and $K$ is a proper cone in $\mathbb{R}^q$:
\begin{align*}\tag{\textsf{CP-STD}}
\min c \cdot x~~\mbox{s.t.}~~Ax = b~~\mbox{and}~~ x \succeq_K 0.
\end{align*}
The above CP has two constraints. The first one, $Ax = b$, is called \emph{non-conic constraint}. The second one, $x \succeq_K 0$ or equivalently $x\in K$, is called \emph{conic constraint}.

LP is a special case of CP, in which $K$ is the non-negative orthant. SDP is a special case of CP, in which $K$ is the set of SPSD matrices.

As in the cases of LP and SDP, there is a dual for CP too, which is also a CP. Before describing the dual, we note that the non-conic constraint $Ax = b$ can be broken into $\ell$ equality constraints $A_h x = b_h$, where $A_h$ is the $h$-th row of the matrix $A$ and $b_h$ is the $h$-th entry of the vector $b$. Each such equality constraint in the primal will associate to one distinct real variable in the dual, but the conic constraint will not associate to any dual variable. This is important since for our problem, we will introduce a CP with exponentially many variables but only \polymn ~equality constraints.
Then its dual will have only \polymn ~dual variables, which is a necessary feature for using ellipsoid algorithm to solve it in \polymn-time.

The dual of \textsf{CP-STD}, in an \emph{inequality form} of CP, is (see \cite[Example 5.12]{BoydV04})
\begin{align*}\tag{\textsf{CP-DUAL-INEQ}}
\max b \cdot y~~\mbox{s.t.}~~c \succeq_{\Ks} A^\intercal y.
\end{align*}
We may solve \textsf{CP-STD} by solving its dual \textsf{CP-DUAL-INEQ} if strong duality holds between them. While strong duality always holds for LP, it may not hold for SDP and CP. The standard method to determine strong duality of CP is to check that \emph{Slater's condition} holds, i.e., there exists an $x$ such that $Ax=b$ and $x\succ_K 0$.

\subsection{Algorithm}\label{subsect:CP-algorithm}

Halperin~\cite{Halperin02} designed an SDP and a rounding scheme for WIS
with approximation guarantee $\bigO\left((\Delta \log\log \Delta)/\log \Delta\right)$.
We conglomerate his SDP with \textsf{LPR-NC} for our problem,
which is equivalent to solving the discrete program \textsf{ICP-C} below.

As the constraint \eqref{eq:bidder-conflict-constraint} involves a product of variables, an LP relaxation is not admissible.
As LP is a subclass of SDP, one might think that an SDP relaxation suffices.
However, this is not true. If we use a ``fully'' SDP relaxation,
each constraint $\xiS\leq 1$ will be converted to a non-conic constraint in the SDP relaxation.
This will introduce exponentially many dual variables, prohibiting an ellipsoid algorithm on its dual to run in poly-time.

Thus, we relax to \textsf{CPR-C}, a ``mixture'' of LP and SDP; note that in \textsf{CPR-C}, $w_0,w_i \in \mathbb{R}^{n+1}$.
\textsf{CPR-C} is a CP. In Appendix~\ref{sect:CP-appendix},
we show that strong duality holds between \textsf{CPR-C} and its dual,
and we can solve the dual in \polymn-time using the ellipsoid algorithm,
assuming that we have a demand oracle for each bidder.
We then round the fractional solution of \textsf{CPR-C} as in Algorithm~\ref{alg:CP}.

\smallskip

\noindent\begin{minipage}{0.50\linewidth}
{
\small{
\begin{align}
\mathrlap{\text{\textsf{(ICP-C)}}}\nonumber\\
\mathrlap{\max \sum_{i\in B} \sum_{S\neq\emptyset} v_i(S) \cdot \xiS}\nonumber\\
\mathrlap{\text{subject to}}\nonumber\\
& \sum_{S\neq\emptyset} \xiS \leq 1, & & \forall i\in B \nonumber\\
& \sum_{S\ni k} \sum_{i\in B} \xiS \leq 1, & & \forall k\in I \nonumber\\
& \frac{1 + \vi}{2} = \sum_{S\neq\emptyset} \xiS, & & \forall i\in B \nonumber\\
& (1 + \vi) (1 + \vj) = 0, & & \forall (i,j)\in E\label{eq:bidder-conflict-constraint}\\
& \vi \in \pm 1, & & \forall i\in B \nonumber\\
& \xiS \in \{0,1\}, & & \forall i\in B,S\subseteq I. \nonumber
\end{align}}}

\vup\vup\vup\vup\vup\vup
\end{minipage}
\begin{minipage}{.50\linewidth}
{
\small{
\begin{align}
\mathrlap{\text{\textsf{(CPR-C)}}}\nonumber\\
\mathrlap{\max ~Z := \sum_{i\in B} \sum_{S\neq\emptyset} v_i(S) \cdot \xiS}\nonumber\\
\mathrlap{\text{subject to}}\nonumber\\
& \sum_{S\neq\emptyset} \xiS \leq 1\nonumber\\
& \sum_{S\ni k} \sum_{i\in B} \xiS \leq 1\nonumber\\
& \frac{1 + w_0 \cdot \vi}{2} = \sum_{S\neq\emptyset} \xiS\label{eq:CPR-sum-ij}\\
& (w_0 + \vi) \cdot (w_0 + \vj) = 0\label{eq:CPR-bidder-item-conflicts}\\
& \|w_0\| = \|\vi\| = 1.\label{eq:CPR-norm}\\
& \xiS \geq 0\nonumber.
\end{align}}}

\vup\vup\vup\vup\vup\vup
\end{minipage}

{\LinesNotNumbered
\begin{algorithm}[t]
\begin{algorithmic}
\State \nl Solve \textsf{CPR-C} to obtain the solution $(Z^*,\{x^*\},\{w^*\})$.
\State \nl Set $\tau\gets \frac{3\log\log \Delta}{4\log \Delta}$, which is less than $1/2$.
Partition the bidders into three sets $B_0,B_1,B_2$:
$B_0 = \left\{i\,|\,0\leq 1 + w_0^* \cdot \vi^* \leq 2\tau\right\}$,
$B_1 = \left\{i\,|\,2\tau < 1 + w_0^* \cdot \vi^* \leq 1 \right\}$,
$B_2 = \left\{i\,|\,1 < 1 + w_0^* \cdot \vi^* \leq 2 \right\}$.
\State\lnl{CP-line:B2}Let $J_2 = B_2$. $\mathcal{A}_2\gets \FCRA(J_2,\hatx(J_2))$.
\State \lnl{CP-line:B1} For the bidders in $B_1$, do as follows:
\begin{itemize}[leftmargin=-0.1em]
\item Project all vectors in $\{\vi^*\,|\,i\in B_1\}$ to $(w_0^*)^\perp$, the space orthogonal to $w_0^*$,
then normalize them. Let $\{\vi '\}$ denote the projected normal vectors.
Note that $(w_0^*)^\perp$ has dimension $n$, so we can treat each $\vi '$ as an $n$-dimensional vector.
\item Choose a random $n$ dimensional vector $r = (r_1,r_2,\cdots,r_n)$,
where each $r_i$ follows the standard normal distribution 
with density function $\phi(x) = \frac{1}{\sqrt{2\pi}} e^{-x^2/2}$.
\item Let $\gamma := (1-2\tau)/(2-2\tau)$.
Let $B_1 ' := \left\{i\in B_1\,|\,\vi '\cdot r \geq \sqrt{\frac{2\gamma}{1-\gamma}\log \Delta}\right\}$.
\item Let $J_1 := B_1 ' \setminus \left\{i\in B_1 '\,|\,\exists j\in B_1 '~\mbox{such that}~(i,j)\in E\right\}$.
$\mathcal{A}_1\gets \FCRA(J_1,\hatx(J_1))$.
\end{itemize}
\State \lnl{CP-line:B0} For the bidders in $B_0$, do as follows:
\begin{itemize}[leftmargin=-0.1em]
\item Let $\{\qiS\}_{i\in B_0,S\subseteq I}$ denote the following distribution:
$\forall S\neq\emptyset$, $\qiS = \frac{\xiS^*}{2\tau\Delta}$, and $\qiE = 1 - \sum_{S\neq \emptyset} \qiS$.
\item $\{T_i\}_{i\in B_0}\gets \FCRA\left(B_0,\{\qiS\}_{i\in B_0,S\subseteq I}\right)$.
\item (Conflict handling.) Let $\mathcal{A}_0$ denote the following allocation:
for each bidder\linebreak $i\in B_0$, if there exists another bidder $j$ such that $(i,j)\in E$ and $T_j\neq\emptyset$,
bidder $i$ gets nothing in $\mathcal{A}_0$; otherwise bidder $i$ gets $T_i$ in $\mathcal{A}_0$.
\end{itemize}
\State \lnl{CP-line:conclude} Return the best allocation among $\mathcal{A}_0,~\mathcal{A}_1,~\mathcal{A}_2$.
\end{algorithmic}
\caption{Approximation Algorithm via Cone Program Relaxation.}
\label{alg:CP}
\end{algorithm}}

\subsection{Intuitions on the Algorithm}\label{subsect:CP-intuition}

Let $\left(Z^*,\{x^*\},\{w^*\}\right)$ be the solution to \textsf{CPR-C}.
Fo any $B'\subseteq B$, let $Z^*(B') := \sum_{i\in B'} \sum_{S\neq\emptyset} v_i(S)\cdot \xiS^*$.

We partition the bidders according to the values of $1 + w_0^* \cdot \vi^*$ into three sets $B_0$, $B_1$ and $B_2$.
Items are allocated to one of the sets; the best one is chosen.
The methods of allocating items to $B_2$ and $B_1$ (Steps \ref{CP-line:B2} and \ref{CP-line:B1})
are well motivated by Halperin's algorithm -- first selecting a ``good'' independent subset of bidders from them,
and then apply FCRA for conflict-free CA; we call this ``IS-then-FCRA''.
In Section \ref{subsect:CP-analysis}, we prove that $\mathcal{A}_2$ and $\mathcal{A}_1$
attain expected social welfares of at least
$\left(1-\frac{1}{e}\right) Z^*(B_2)$ and $\Omega\left(\frac{\log \Delta}{\Delta \log\log \Delta}\right)\cdot Z^*(B_1)$ respectively.

For $B_0$, we face two difficulties which force us to use an approach quite different from Halperin's.
Firstly, Halperin's algorithm is for undirected graph while in our application the graph is directed.
Secondly, we notice that the ``IS-then-FCRA'' approach will not work for $B_0$,
and we ought to do the opposite -- first apply FCRA by ignoring conflicts (see the next paragraph),
and then resolve any remaining conflicts.
These force us to have an analysis for $B_0$ quite different from Halperin's one.

We provide more intuitions for $B_0$.
The bidders in $B_0$ have low values of $\frac{1 + w_0^* \cdot \vi^*}{2} = \sum_{S\neq\emptyset} \xiS^*$.
The values $\xiS^*$ are typically viewed as probability densities.
Low values of $\frac{1 + w_0^* \cdot \vi^*}{2}$ allow room to ``expand'' these densities by a factor of $1/\tau$,
where $\tau<\frac{1}{2}$.
However, to handle conflicts, we ought to ``dwell'' these densities by a factor of $1/(2\Delta)$ afterwards.
Then we apply FCRA with the ``expanded then dwelled'' densities to obtain
a sufficiently good allocation to $B_0$.
These will allow us to show that $\mathcal{A}_0$ attains an expected social welfare of
$\Omega\left(\frac{\log \Delta}{\Delta \log\log \Delta}\right)\cdot Z^*\left(B_0\right)$.

Finally, note that $Z^*(B_0) + Z^*(B_1) + Z^*(B_2) = Z^*$, so the best among $\mathcal{A}_0,~\mathcal{A}_1,~\mathcal{A}_2$
attains an expected social welfare of $\Omega\left(\frac{\log \Delta}{\Delta \log\log \Delta}\right)\cdot Z^*$.

\subsection{Analysis}\label{subsect:CP-analysis}

We need the following notation.
Recall that for any set $B'\subseteq B$, $Z^*(B') := \sum_{i\in B'} \sum_{S\neq\emptyset} v_i(S)\cdot \xiS^*$.
Note that $Z^*(B_0) + Z^*(B_1) + Z^*(B_2) = Z^*$;
let $L^*(B')$ denote the optimal objective value of $\LPR (B')$;
let $x^*(B')$ denote the vector $x^*$ restricted to bidders in $B'$.

\medskip

\noindent\underline{Analysis on Step \ref{CP-line:B2}.}
Constraints \eqref{eq:CPR-bidder-item-conflicts} guarantee that $J_2 = B_2$ is an independent set.
Since $x^*(J_2)$ is a feasible point of $\LPR(J_2)$, $L^*(J_2)\geq Z^*(J_2) = Z^*(B_2)$.
FCRA gives an allocation which is at least
$\left(1-\frac{1}{e}\right) L^*(J_2)\geq \left(1-\frac{1}{e}\right) Z^*(B_2)$ in expectation.

\medskip

\noindent\underline{Analysis on Step \ref{CP-line:B1}.}
We note that the selection of independent set $J_1$ is identical to the corresponding part in Halperin's algorithm, modulo that since we are dealing with a directed graph, we can remove fewer bidders from $B_1'$.\footnote{Halperin's analysis is for undirected graphs, but his proof can be reused for directed graphs with little modification. In \cite[Lemma 5.2]{Halperin02}, if there is an edge between vertices in $B_1'$, both vertices of the edge are removed from $B_1'$. For directed graphs it suffices to remove the outgoing vertex only, so the bound provided in the lemma is also applicable.}
Thus, we can follow closely to Halperin's analysis to show that
$\Ex{Z^*(J_1)} = \Omega\left(\frac{\log \Delta}{\Delta \log\log \Delta}\right)\cdot Z^*(B_1).$

Since $x^*(J_1)$ is a feasible point of $\LPR(J_1)$, $L^*(J_1)\geq Z^*(J_1)$.
FCRA gives an allocation which is at least
$\left(1-\frac{1}{e}\right) L^*(J_1) = \Omega\left(\frac{\log \Delta}{\Delta \log\log \Delta}\right)\cdot Z^*(B_1)$ in expectation.

\medskip

\noindent\underline{Analysis on Step \ref{CP-line:B0}.}
Observe that $2\tau\Delta > 1$ for sufficiently large $\Delta$,
so the vector $q$, which collects $\{\qiS\}_{i\in B_0,S\subseteq I}$, is a feasible point of $\LPR(B_0)$.

For the analysis of this step, we need to unwind FCRA.
Taking the feasible point $q$ as input, the algorithm first selects a random set $S_i$ for each bidder $i$ as follows:
a non-empty set $S$ is selected with probability $\qiS$, and the empty set is selected with probability $1-\sum_{S\neq\emptyset} \qiS$.
Note that the random sets $S_1,S_2,\cdots,S_n$ may not be disjoint, so the algorithm next carries on a \emph{resolution scheme}
to randomly generate disjoint sets $T_1,T_2,\cdots,T_n$, which are the sets stated in Step \ref{CP-line:B0},
while for all $i$, $T_i\subseteq S_i$.

By Lemma \ref{lem:FCRA}, $\Ex{v_i(T_i)}$, the expected welfare of bidder $i$ (modulo conflicts), is at least
$\left(1-\frac{1}{e}\right)\frac{Z^*\left(\{i\}\right)}{2\tau \Delta}$.

To handle conflicts, the algorithm resets the allocation of some bidders to the empty set.
We will show that for each bidder $i\in B_0$,
at least half of his expected welfare (modulo conflicts) is retained after conflict handling.

For every $i\in B_0$, let $F_i$ be the event: $\forall j$ with $(i,j)\in E$, $S_j=\emptyset$.
Then $\overline{F_i}$ is the event: $\exists j$ such that $(i,j)\in E$ and $S_j\neq\emptyset$.
We note that before conflict handling, for all $i\in B_0$, $\Ex{v_i(T_i)\,|\,F_i} \geq \Ex{v_i(T_i)\,|\,\overline{F_i}}$.
We will prove the above inequality formally in Lemma \ref{lem:more-with-less-competition} below,
but it is indeed intuitive in the following sense:
$S_i$ is the set of items the bidder $i$ competes for during the resolution scheme,
thus the above inequality depicts that bidder $i$ gets more when facing less competition from bidders he conflicts with.

Note that $\Ex{v_i(T_i)} = \Ex{v_i(T_i)\,|\,F_i} \cdot \Pr{F_i} + 
\Ex{v_i(T_i)\,|\,\overline{F_i}} \cdot \Pr{\overline{F_i}}$,
i.e., $\Ex{v_i(T_i)}$ is a weighted average of the two conditional expectations.
Since the first conditional expectation is larger than the second one,
$$\Ex{v_i(T_i)\,|\,F_i} \geq \Ex{v_i(T_i)} \geq \left(1-\frac{1}{e}\right)\frac{Z^*\left(\{i\}\right)}{2\tau \Delta}.$$

Next, note that $\Pr{F_i} = 1 - \Pr{\overline{F_i}}
= 1- \Pr{\cup_{j:(i,j)\in E} \left(S_j\neq\emptyset\right)}$ is at least
\begin{align*}
1-\sum_{j:(i,j)\in E} \Pr{S_j\neq\emptyset} &= 1-\sum_{j:(i,j)\in E} \frac{1}{2\tau\Delta}\sum_{S\neq\emptyset} x^*_{j,S} \geq 1-\sum_{j:(i,j)\in E} \frac{1}{2\tau\Delta}\cdot \tau\qquad(j\in B_0)\\
&~~~~\geq 1/2.\qquad(i~\mbox{conflicts with at most}~\Delta~\mbox{bidders})
\end{align*}

Bidder $i$'s allocation is reset during conflict handling only if $\overline{F_i}$ holds.
By the last two paragraphs, the expected welfare of bidder $i$ after conflict handling is at least
$$\Ex{v_i(T_i)\,|\,F_i}\cdot \Pr{F_i}
\geq \left(1-\frac{1}{e}\right)\frac{Z^*\left(\{i\}\right)}{2\tau \Delta} \cdot \frac{1}{2}
= \Omega\left(\frac{\log \Delta}{\Delta \log\log \Delta}\right)\cdot Z^*\left(\{i\}\right).$$
Then the expected social welfare is at least
$\sum_{i\in B_0} \Omega\left(\frac{\log \Delta}{\Delta \log\log \Delta}\right)\cdot Z^*\left(\{i\}\right) = \Omega\left(\frac{\log \Delta}{\Delta \log\log \Delta}\right)\cdot Z^*\left(B_0\right)$.

\medskip

\noindent\underline{Analysis on Step \ref{CP-line:conclude}.} The final step is to choose the best allocation among $\mathcal{A}_0,~\mathcal{A}_1,~\mathcal{A}_2$,
which, by the analyses of the previous three steps, is at least
$$\frac{1}{3}\left[\left(1-\frac{1}{e}\right) Z^*(B_2) + \Omega\left(\frac{\log \Delta}{\Delta \log\log \Delta}\right)\cdot \left(Z^*(B_1) + Z^*(B_0)\right)\right] = \Omega\left(\frac{\log \Delta}{\Delta \log\log \Delta}\right)\cdot Z^*.$$

This concludes the analysis of the algorithm.

\medskip

\begin{lemma}\label{lem:more-with-less-competition}
Let $S_i,S_j,T_i$ be as defined above.
For every $i\in B_0$, let $F_i$ be the event: $\forall j$ with $(i,j)\in E$, $S_j=\emptyset$.
Then $\Ex{v_i(T_i)\,|\,F_i} \geq \Ex{v_i(T_i)\,|\,\overline{F_i}}$.
\end{lemma}
\begin{pf}
Let $C_i$ be the set of bidders $j$ with $(i,j)\in E$.
Fix $S_j$ of bidders $j\in B_0\setminus C_i$.
For an item $k\in S_i$ and every $i'\in B_0$, let $p_{i'}(k) := \sum_{S\ni k} x_{i',S}^*$.
Let $A(k) := \left\{i'\,|\,S_{i'} \ni k\right\}$.

In FCRA (see \cite[Section 1.2]{FeigeV10}),
if $|A(k)|=1$, i.e.~$A(k)=\{i\}$, then $\Pr{k\in T_i} = 1$;
if $|A(k)|>1$, then 
$$\Pr{k\in T_i} = \frac{1}{\sum_{i'\in B_0} p_{i'}(k)}\left(\sum_{i'\in A(k)\setminus \{i\}} \frac{p_{i'}(k)}{|A(k)|-1} + \sum_{i'\notin A(k)} \frac{p_{i'}(k)}{|A(k)|}\right).$$

Recall that we are fixing $S_j$ of $j\in B_0\setminus C_i$.
If $F_i$ holds, then $\forall j\in C_i$, $j\notin A(k)$.
However, if $\overline{F_i}$ holds, some bidders in $C_i$ may get into $A(k)$, i.e.,
$$(|A(k)|~\mbox{when}~F_i~\mbox{holds}) \leq (|A(k)|~\mbox{when}~\overline{F_i}~\mbox{holds}),$$
no matter what $S_j$ the bidders $j$ in $C_i$ choose. Then it is easy to see that
$$\Pr{k\in T_i\,|\,F_i} \geq \Pr{k\in T_i\,|\,\overline{F_i}}.$$

As each item is allocated independently,
and each item in $S_i$ is allocated to bidder $i$ with higher probability when $F_i$ holds,
the lemma follows.
\end{pf}

\subsection{Discussion}\label{subsect:CP-discussion}

We obtain the following proposition as a notable special case.
\begin{proposition}\label{coro:WIS-directed}
There is a poly-time $\bigO\left((\Delta \log\log \Delta)/\log \Delta\right)$-approximation algorithm for
the WIS problem in a directed graph $G$ with out-degree at most $\Delta$.
\end{proposition}
Its proof is via a simple reduction from WIS to our problem.
Consider a CA in which the bidder conflicts are represented by the directed graph $G$.
Each bidder wants one distinct item, and his valuation of the item is the weight of his corresponding vertex in the WIS problem.

\medskip

By Theorem \ref{thm:randmain2}, we have an $\bigO\left(\min\{\Delta,\Delta_I\}\right)$-approximation algorithm
for CA with bidder and item conflicts, in which bidders have FSA valuations.
An interesting open problem is whether it is possible to improve the approximation guarantee to $o\left(\min\{\Delta,\Delta_I\}\right)$.
We note that if each bidder has linear valuation, the problem reduces to the WIS problem
in the \emph{tensor product} of the graphs $G$ and $G_I$, which might be a problem of independent interest.

Motivated by our CP relaxation \textsf{CPR-C} for CA with bidder conflicts,
a valid CP relaxation for CA with bidder and item conflicts can be obtained
by replacing \eqref{eq:CPR-sum-ij}--\eqref{eq:CPR-norm} in \textsf{CPR-C} with
\begin{align*}
& \frac{1 + w_0 \cdot w_{ik}}{2} = \sum_{S\ni k} \xiS, & \forall i\in B,k\in I\tag{\ref{eq:CPR-sum-ij}'}\\
& (w_0 + w_{i_1k_1}) \cdot (w_0 + w_{i_2k_2}) = 0, & \forall (i_1,i_2)\in E,~(k_1,k_2)\in E_I\tag{\ref{eq:CPR-bidder-item-conflicts}'}\\
& \|w_0\| = \|w_{ik}\| = 1, & \forall i\in B,k\in I.\tag{\ref{eq:CPR-norm}'}
\end{align*}
However, we do not see a good rounding algorithm for this relaxation.

%% file: SubLinSponsoredSearch.tex

\section{Sponsored Search with Limited Number of Slots}
\label{sec:adwords}

In this section we consider sponsored search with bidder conflicts.  Some of our results extend to ordered conflicts and more general graph-based slot conflicts. In light of the application, we concentrate on the case with a small number $m$ of slots. Note that a trivial enumeration solves the problem in time $\bigO(n^m)$. Moreover, it is unlikely that significantly faster algorithms exist that solve the problem exactly, even for $m \le \log n$; it is \classWOne-hard to decide Log-Independent-Set, i.e., given $k \le \log n$, deciding if $G$ has an independent set of size at least $k$ cannot be done in time $f(k) \cdot n^c$ for constant $c$ unless \classFPT\ = \classWOne~\cite{DowneyF99}. Thus, we present two approximation algorithms. The first one uses semi-definite programming and has polynomial running time for $m \in \bigO(\log n)$. The second one is a partial enumeration approach and runs in polynomial time if $m \in \bigO((\log n)/(\log \max(\Delta+1,\log n)))$.

\subsection{Sponsored Search via Semidefinite Programming}
%
We study sponsored search with bidder conflicts and $m \in \bigO(\log{n})$.
We assume for simplicity that $n\geq m\geq 6$.
If $m>n$, we could add $(m-n)$ dummy bidders with valuation zero.
We assume consistent tie-breaking among bidders with the same valuation.
Recall that in this setting bidders have unit demands, and thus we can represent an allocation $S$ of slots to bidders by a matching $M_S$ in a bipartite bidder-slot-graph. We define $v_i(M_S)=v_i(S)$ for all $i\in B$. We call a matching $M_S$ \emph{conflict-free} if $D_i\cap S_i=\emptyset$ for all $i\in B$. Note that for every matching there exists a conflict-free matching with the same social welfare; we simply unassign all the slots in $\bigcup_{i\in B}D_i\cap S_i$. Furthermore, we define the expected social welfare $\swe(M):=\mathbf{E}[\sum_{i\in B}v_i(M)]$ for a (randomized) matching $M$. In the following, we also use the notation $\rd :=\sqrt{\log\log\Delta/\log\Delta}$.

\begin{algorithm}[t]
Assign all bidders in $B$
independently with probability $1/2$ to set $B_1$ and set $B_2\gets B\setminus B_1$.
With  $v_1 \ge v_2 \ge \dots \ge v_n$ and $h = |B_1|$ define the functions $\orderone:[h] \rightarrow [n]$ and $\ordertwo:[n-h] \rightarrow [n]$ such that
$B_1 = \{\orderone(1), \dots ,\orderone(h)\}$
and ${\orderone(j)} < {\orderone(j+1)}$ for $j \in [h-1]$  and
$B_2 = \{\ordertwo(1), \dots ,\ordertwo(n-h)\}$ and
${\ordertwo(j)} < {\ordertwo(j+1)}$ for $j \in [n-h-1]$\;
Set $q\gets 1$ with probability $\frac{1}{2}$ and set $q\gets 2$ otherwise\;
\lIf{$n-h\ge \lceil \frac{m}{4}\rceil+1$}{
$t \gets \ordertwo(\lceil\frac{m}{4}\rceil+1)$\ 
\textbf{else}\ $t\gets \infty$ and $v_t\gets 0$\;}
\eIf{$q=1$}
{
Set $r_1\gets v_t$;
Set $B^1_1 \gets \{\orderone(j)| j \in [h] $ and $\orderone(j) < t\}$\;
\lIf{$t\le m+1$}{
set $\mathcal{A}$ to the set of all subsets of $B^1_1$\label{ln:t1good}
\textbf{else}\ $\mathcal{A}\gets \emptyset$}
}
{
Set  $r_2\gets v_t \cdot\frac{1}{8}\rd$;
Set $B^2_1 \gets \{\orderone(j)| j \in [h] $ and $v_{\orderone(j)} \ge r_2\}$\;


Set $J\gets$ (unweighted) independent set in $B^2_1$ computed by using the WIS algorithm (Proposition~\ref{coro:WIS-directed}) giving bidders in $B^2_1$ in random order and with equal weights; $\mathcal{A}\gets \{J\}$\;
}
Add $m$ bidders without conflicts and with valuation $r_q$ to $B$ and each set in $\mathcal{A}$\;
For each set $A\in \mathcal{A}$ let $\mathcal{M}(A)$ define all the conflict-free matchings of bidders in $A$ to slots; define $\mathcal{M}=\bigcup_{A\in\mathcal{A}}\mathcal{M}(A)$\;
Select allocation $M'\in\arg\max_{M\in \mathcal{M}} \sum_{i\in B} v_i(M)$\label{ln:comp1}\;
Every real-bidder $a$ in $B$ pays $p_a\gets\max_{M\in \mathcal{M}} \sum_{i\in B\setminus\{a\}}( v_i(M)-v_i(M'))$\label{ln:comp2}\;
\caption{Sponsored search auction with conflicts}
\label{alg:sponsored}
\end{algorithm}

The mechanism is presented in Algorithm~\ref{alg:sponsored} and its approximation guarantee is analyzed in Lemma~\ref{thm:ssapprox}.
Let $t$, $r_1$, $r_2$, $B^1_1$, and $B^2_1$ be defined as in Algorithm~\ref{alg:sponsored}.
We show that if the optimal conflict-free assignment of bidders to slots $\OPT$ was restricted to a random subset $\OPT''$ of the $t-1$ most valuable edges, where each of those edges is picked with probability $1/2$, then $\swe(\OPT'')\ge \swe(\OPT)/16$. Thus, it suffices to compare the performance of a mechanism with $\OPT''$. We run two different mechanisms, $\mathsf{ALG}_1$ and $\mathsf{ALG}_2$, each with probability $1/2$, and receive at least $1/2$ of the maximum of their social welfares $\sw_1$ and $\sw_2$, respectively.

If $\mathsf{ALG}_1$ performs very well, i.e., if $\swe_1 > \swe(\OPT'')/(\Delta \rd )$, we achieve the result promised in Lemma~\ref{thm:ssapprox}. Mechanism $\mathsf{ALG}_1$ tries out all possibilities to find the best non-conflicting matching for bidders in $B^1_1$.
If $\mathsf{ALG}_1$ does not perform very well, we can show that $\OPT''$ must get at least a quarter of its social welfare from bidders in $B^2_1\setminus B^1_1$. In this case, we build an (unweighted) independent set $J$ of all bidders in $B^2_1$ using the WIS algorithm described in Proposition~\ref{coro:WIS-directed}, which guarantees that the number of bidders in $J$ is at least an $\mathcal{O}(1/(\Delta \rd^2))$-fraction of the optimal number for bidders in $B^2_1\setminus B^1_1$. As in $\OPT''$ every bidder in $B^2_1\setminus B^1_1$ contributes at most with valuation $r_1$ to $\swe(\OPT'')$ and in $\mathsf{ALG}_2$ every bidder in $J$ contributes at least with valuation $r_2$ to $\sw_2$, the overall approximation ratio of $\mathsf{ALG}_2$ is $\mathcal{O}(\Delta \rd^2\cdot r_1/r_2 )=\mathcal{O}(\Delta \rd )$. 
\begin{lemma}
The matching $M'$ computed in Algorithm~\ref{alg:sponsored} is in expectation an\\
$\mathcal{O}(\Delta \sqrt{\log\log\Delta/\log\Delta})$-approximation of the optimal social-welfare.
\label{thm:ssapprox}
\end{lemma}

To proceed, we need the lemma below.

\begin{lemma}
It holds that $\Pr{t \le m+1}\ge 3/4$.
\label{lem:boundprob}
\end{lemma}
\begin{pf}
Note that $t\le m+1$ if and only if $\left|B_2\cap [m+1]\right| \geq \lceil \frac{m}{4} \rceil + 1$.
This happens with probability $1 - \frac{1}{2^{m+1}} \sum_{\ell=0}^{\lceil \frac{m}{4} \rceil} \binom{m+1}{\ell}$,
which is at least $3/4$ when $m\geq 6$.
\end{pf}

\begin{pfof}{Lemma~\ref{thm:ssapprox}}
Assume that the social-welfare-maximizing conflict-free\linebreak matching of bidders in $B$ to slots is given by $\OPT$.  The valuation of the dummy bidders will not be considered in the social welfare as they were only included to guarantee truthfulness.

We first analyze the random partition of $B$ into $B_1$ and $B_2$ by the mechanism. Let $m^*$ be the number of edges in $\OPT$ and let us denote those edges by $(i(1),{j}(1)),\dots,(i(m^*),{j}(m^*))$ such that they are ordered by their value, i.e., $v_{i(1)}\cdot \alpha_{{j}(1)}\ge\cdots\ge v_{i(m^*)}\cdot \alpha_{{j}(m^*)}$.
Let $\OPT'$ be the random subset of $\OPT$ where all the edges but the $t-1$ most valuable ones are discarded, i.e., $\OPT'=\{(i(1),{j}(1)),\dots,(i(t-1),{j}(t-1))\}$. Furthermore, let $\OPT''$ be the random subset of $\OPT'$ where (1) all the edges that contain bidders in $B_2$ are discarded and (2) if $t>m+1$ all edges are discarded.
We will show that $\swe(\OPT)\le 16\cdot\swe(\OPT'')$.
Since, $B_2\subseteq B$ it holds that $t\ge \lceil m/4\rceil+1\ge m/4+1$, and thus, it follows by $m^*\le m$ that
$$\frac{\swe(\OPT)}{\swe(\OPT')}=\frac{\sum_{s=1}^{t-1} v_{i(s)}\cdot \alpha_{{j}(s)}+\sum_{s=t}^{m^*} v_{i(s)}\cdot \alpha_{{j}(s)}}{\sum_{s=1}^{t-1} v_{i(s)}\cdot \alpha_{{j}(s)}} \le 1+\frac{\sum_{s=t}^{m^*} v_{i(t)}\cdot \alpha_{{j}(t)}}{\sum_{s=1}^{t-1} v_{i(t)}\cdot \alpha_{{j}(t)}}=\frac{m^*}{t-1}\le 4\,.$$
Now, for all $i\in B$ let $E_i$ be the event that bidder $i$ is not in $B_1$ and let $T$ be the event that $t> m+1$. By Lemma~\ref{lem:boundprob}, it holds for each bidder $i\in B$ that $\Pr{E_i\cup T}\le \Pr{E_i}+\Pr{T}\le 1/2+1/4=3/4$. Thus, $\swe(\OPT'')=\sum_{s=1}^{t-1} v_{i(s)}\cdot \alpha_{{j}(s)}\cdot(1-\Pr{E_{i(s)}\cup T})\ge(1/4)\cdot \swe(\OPT')$.
It follows that $\swe(\OPT'')\ge (1/16)\cdot \swe(\OPT)$.

We will now compare the outcome $M'$ of the mechanism with $\OPT''$.
Let $M_1$ or $M_2$ be the matching computed by the mechanism under the condition $q=1$ or $q=2$, respectively. It holds that
$2\cdot\swe(M')\ge \max\{\swe(M_1),\swe(M_2)\}$. Then the following claim completes the proof.

\begin{claim}
For some constant $c>1$ it holds that
\begin{equation}
\label{eq:case1}
c\cdot \Delta\rd \cdot\max\{\swe(M_1),\swe(M_2)\}\ge \swe(\OPT'').
\end{equation}
\end{claim}

\begin{pf}
Notice that if $\swe(\OPT'')< \max\{4,\Delta \rd \} \cdot \swe(M_1)$
then (\ref{eq:case1}) is satisfied. 
Thus, we assume that $\swe(\OPT'')\ge \max\{4,\Delta \rd \} \cdot \swe(M_1)$. Moreover, we assume that $t\le m+1$, as otherwise, $\swe(\OPT'')=0$.

Next, we define by $\swu$ the optimal social welfare for bidders in $B^1_1$ when bidder conflicts are ignored. Furthermore, note that Theorem~\ref{thm:detalg} implies that $4\Delta\cdot \swe(M_1)\ge\swu$.
Thus,
\begin{equation}
\label{eq:case2}
\swe(\OPT'')\ge\Delta \rd \cdot \swe(M_1)\ge \Delta \rd \cdot \swu/(4\Delta)\ge (\rd /4)\cdot\swu
\end{equation}
Let us now partition the matching $\OPT''$ into $\OPT_1$ that contains the edges to bidders in $B^1_1$, $\OPT_2$ that contains the edges to bidders in $B^2_1\setminus B^1_1$, and $\OPT_3:=\OPT''\setminus(\OPT_1\cup\OPT_2)$. Thus, $\swe(\OPT'')=\swe(\OPT_1)+\swe(\OPT_2)+\swe(\OPT_3)$.

As the matching $\OPT_1$ is considered when computing $M_1$, $\swe(M_1)\ge\swe(\OPT_1)$, and thus, by the assumption taken above holds that
$$\swe(\OPT_1)\le \swe(M_1)\le \swe(\OPT'')/\max\{4,\Delta \rd \}\le\swe(\OPT'')/4.$$

Furthermore, $\swe(\OPT_3)\le \swe(\OPT'')/2$, as otherwise, 
\[\swe(\OPT'')<2\cdot\swe(\OPT_3)< 2\cdot r_2\cdot \sum_{{j}=1}^{t-1} \alpha_{j}=2\cdot \frac{1}{8}\cdot \rd \cdot v_{{t}}\cdot \sum_{{j}=1}^{t-1} \alpha_{j}\le \frac{\rd}{4} \swu\,,\]
which contradicts (\ref{eq:case2}). Hence, $\swe(\OPT_2)\ge\swe(\OPT'')/4$. It follows that
\begin{multline}
\frac{\swe(\OPT_2)}{\swe(M_2)}=\frac{\sum_{({i},{j})\in\OPT_2}v_{i}\cdot \alpha_{j}}{\sum_{({i},{j})\in M_2}v_{i}\cdot \alpha_{j}}\le \frac{r_1\cdot \sum_{{j}=1}^{|\OPT_2|}\alpha_{j}}{r_2\cdot\sum_{{j}=1}^{|M_2|}\alpha_{j}}\\
\le \frac{8}{\rd }\cdot\bigg(1+\frac{\sum_{{j}=|M_2|+1}^{|\OPT_2|}\alpha_{|M_2|}}{\sum_{{j}=1}^{|M_2|}\alpha_{|M_2|}}\bigg)
= \frac{8}{\rd }\cdot\frac{|\OPT_2|}{|M_2|}\\
\le\frac{8}{\rd }\cdot c'\cdot(\Delta\cdot \rd^{2})=8c'\cdot \Delta\cdot \rd ,
\end{multline}
where $c'\Delta \rd^{2}$ is the approximation factor of Proposition~\ref{coro:WIS-directed}.
\end{pf}
\end{pfof}

We show that the mechanism runs in \textsf{poly}$(n,\Delta)$ time for certain restrictions on the number of slots $m$, and it is universally truthful.
The crucial idea for showing truthfulness is to prove that no bidder has an incentive to alter the set of matchings $\mathcal{M}$.
Thus, even though the range of allocations $\mathcal{M}$ depends on the valuations of the bidders, no bidder has an incentive to change it.

\begin{proposition}
If $m\in\mathcal{O}(\log{n})$ the mechanism takes time \textsf{poly}$(n,\Delta)$.
\label{prop:runtime}
\end{proposition}
\begin{pf}
We have to show that line \ref{ln:comp1} and \ref{ln:comp2} can be computed in polynomial time in $n$ and $\Delta$.

We first argue that $|\mathcal{A}|$ is polynomial in $n$. Consider the case where $q=1$. If $t> m+1$ then $\mathcal{A}=\emptyset$; otherwise, $\mathcal{A}=\mathcal{P}(B^1_1)$ and $|B^1_1|<t\le m+1$. Moreover, if $q=2$ then $|\mathcal{A}|=1$. Thus, $|\mathcal{A}|$ is bounded by $2^m$ which is polynomial in $n$.

Next, assume that we are given some $A\in \mathcal{A}$. 
Given $q=1$, we can ignore $A$ if bidders in $A$ have conflicts, because we know that there exists a conflict-free set of bidders in $\mathcal{A}$ that is optimal. Moreover, given $q=2$ we know that the bidders in all sets in $\mathcal{A}$ have no conflicts. Thus, we can assume that the bidders in $A$ are conflict-free. It follows that computing $\arg \max_{M\in \mathcal{M}(A)}\sum_{i\in C} v_i(M)$ can be done in polynomial time in $n$ for all $C\subseteq B$; for all $i\in [m]$ the bidder in $A$ with the $i$-th largest index has to be matched to the $i$-th slot.
\end{pf}

\begin{lemma}
\label{lem:rangefixed}
No bidder has an incentive to report a non-truthful bid that alters~$\mathcal{M}$.
\end{lemma}
\begin{pf}
We can restrict the proof to bidders in $B_1$ as the other bidders always have utility zero. Assume that all bidders bid truthful. In both cases, $q=1$ and $q=2$, bidders not in $B^q_1$ are in no matching in $\mathcal{M}$, and thus, they are not in $M'$ and their utility is zero. However, they have no incentive to increase their bid because there are $m$ competing dummy-bidders that have a valuation that is at least the same as theirs. Thus, if they increase their valuation, their utility cannot increase because they have to pay their externality. Furthermore, in both cases, $q=1$ and $q=2$, bidders in $B^q_1$
have two possibilities: (i) Bidding high enough to stay in $B^q_1$ and (ii) bidding below the value that is necessary for staying in $B^q_1$. In (i), if a bidder bids high enough to stay in $B^q_1$, he cannot affect $B^q_1$. Moreover, he cannot affect the outcome of the WIS algorithm by his bid because we randomized the order of the bidders. Thus, he cannot influence whether he belongs to a subset in $\mathcal{A}$ and, in turn, he cannot influence $\mathcal{M}$.
In (ii), if a bidder bids below the value that is necessary for staying in $B^q_1$, he will receive nothing and has utility zero.
Thus, no bidder has an incentive to change his bid if this alters $\mathcal{M}$.
\end{pf}

\begin{lemma}
The mechanism is universally truthful.
\label{thm:truthful}
\end{lemma}
\begin{pf}
We can assume that all random decision are taken before the bidders report their bids.
We first fix a bidder $a$.
Since by Lemma~\ref{lem:rangefixed} no bidder has an incentive to report a non-truthful bid that changes $\mathcal{M}$,
we can restrict the analysis to bidder $a$'s non-truthful bids that do not change $\mathcal{M}$. Thus, we can  consider $\mathcal{M}$ as fixed. The utility of a bidder $a$ for a matching $M'$ is given by $u_a(M')=v_a(M')-(\max_{M\in \mathcal{M}} \sum_{i\in B\setminus\{a\}}( v_i(M)-v_i(M')))=\sum_{i\in B} v_i(M')-\max_{M\in\mathcal{M}}\sum_{i\in B\setminus \{a\}}v_i(M)$ which is maximized when $a$ bids truthful.
\end{pf}

The following theorem follows from Lemma \ref{thm:ssapprox}, Proposition~\ref{prop:runtime} and Lemma~\ref{thm:truthful}.

\begin{theorem}
\label{thm:ssmainthm}
For sponsored search with bidder conflicts and $m\in\mathcal{O}(\log{n})$, Algorithm~\ref{alg:sponsored} is a universally-truthful mechanism that
attains approximation guarantee of $\mathcal{O}(\Delta \sqrt{\log\log\Delta/\log\Delta})$.
It runs in time \textsf{poly}$(n,\Delta)$.
\end{theorem}

%% file: sponsored_search_limited_number_slots.tex
\subsection{Sponsored Search via Partial Enumeration}
%
We treat a slightly more general \emph{small-supply} case with $m \le n/(\Delta+1)$.
For this case we observe that the problem can be solved optimally in linear time when all bidders $i$ have uniform values $v_i = v$. 
For non-uniform values $v_i$, we will strive for a truthful mechanism that solves the problem approximately
but much faster than the trivial enumeration that solves the problem exactly in $\bigO(n^m)$ time.
Note that there is an $m$-approximation algorithm that assigns slot 1 to the highest bidder,
obtains value $\max_{k,i} \alpha_k \cdot v_i$, and runs in time $\bigO(n)$.
Thus, we obtain the following trade-off.

\begin{theorem}\label{thm:small-supply}
In sponsored search with bidder and slot conflicts, there is a\linebreak universally-truthful mechanism that yields an $\bigO(\log m)$-approximation of social welfare and runs in time $\bigO(n+(m(\Delta+1))^m)$. 
\end{theorem}

We first prove the existence of an approximation algorithm achieving the claimed approximation ratio.

\begin{lemma}\label{prop:small-supply}
In sponsored search with bidder conflicts, there is an $\bigO(\log m)$-approximation algorithm that runs in time $\bigO(n+(m(\Delta+1))^m)$. 
\end{lemma}

\begin{pf}
	The algorithm is extremely simple for uniform values $v_i = v$ for all $i \in B$ if $m\le n/(\Delta+1)$. Initially, every bidder is active. We assign slot 1 to the bidder $i$ with smallest out-degree, label $i$ and its all out-neighbors to be inactive. We repeat this procedure with slots $2,3,\ldots,m$. Since $m \le n/(\Delta+1)$, we will be able to assign all slots in this way.
This yields an optimum solution and takes time $\bigO(n)$.
If the $v_i$ are different, we apply logarithmic scaling. Let $v_{\max} = \max_{i \in B} v_i$. We consider $\lceil \log_2 (2m) \rceil$ classes, where class $k$ contains bidders $i$ with value $v_i \in (v_{\max}/2^k,v_{\max}/2^{k-1}]$. The unclassified bidders have a value which is at most $v_i \le v_{\max}/(2m)$. Thus, by discarding this set of bidders, we discard at most 1/2 of the optimum value. 

For the remaining bidders, we pick $k \in \{1,2,\ldots,\lceil \log_2 (2m)\rceil\}$ uniformly at random and consider $V_k = \{ i \in B \mid v_i > v_{\max}/2^k \}$, the union of all bidders in classes $1,\ldots,k$. Let $n_k = |V_k|$. If $n_k / (\Delta+1) \ge m$, then we can apply the above algorithm for identical values to $V_k$. Otherwise, if $n_k / (\Delta+1) \le m$, then $n_k \le (\Delta+1) m$, and a complete enumeration takes time at most $\bigO((m(\Delta+1))^m)$. In either case, we obtain the optimum for $V_k$ under the assumption that every bidder has value $v_{\max}/{2^k}$, and hence at least half of the value that the optimum gets from bidders in class $k$. In expectation over the random choice of $k$, this shows that we recover an $\bigO(\log m)$-fraction of the optimum.

The highest valuation can be found in time $\bigO(n)$. Computing the threshold and reducing the set of considered bidders can be done in time $\bigO(n)$. Applying the previous algorithm can be done in time $\bigO(n)$, enumeration takes time $\bigO((m(\Delta+1))^m)$.
\end{pf}

Note that for a particular choice of $k$, the algorithm described in the proof of Lemma~\ref{prop:small-supply} is applied in the induced subgraph of $V_k$ and produces an optimum solution under the assumption that all nodes have the same valuation. If this results from the greedy algorithm for the independent set of bidders, it also remains an optimum solution with arbitrary additional slot conflicts. If this results from enumeration, we can apply the enumeration also for additional slot conflicts in the same asymptotic running time. Thus, we obtain the same running time and approximation ratio also for sponsored search with bidder and slot conflicts.

By the sampling arguments in \cite{DobzinskiNS12,HoeferK13} we can turn the algorithm into a universally truthful mechanism
with the same asymptotic running time and approximation ratio. The idea is as follows.
First, choose a random bit $q$.
If $q=0$, partition $B$ into $B_1$ and $B_2$ randomly and set $v_{\max}$ be the highest valuation in $B_1$.
However, we run the algorithm in Proposition \ref{prop:small-supply} on $B_2$ only; 
if bidder $i\in B_2$ gets assigned slot $\ell$ he has to pay $\alpha_\ell\cdot v_{\max}/2^k$.
If $q=1$, we keep the best slot and remove all others, and run a second price auction among all bidders in $B$.
This ensures that the claimed approximation ratio even if there is a dominant bidder,
i.e., a bidder who contributes at least a constant fraction of the optimal social welfare.
This completes the proof of Theorem \ref{thm:small-supply}.

%% file: Rounding-Appendix.tex

\section{Proofs and Results Omitted in Section \ref{sec:randmech}}
\label{app:bounded_item_degree}




\subsection{Derandomization}\label{subsect:sampling-derandomization}



It is crucial for the derandomization to show that
there exists a pairwise independent distribution of subsets of $B$ with probability $1/2^{\lceil \log_2 \Delta\rceil +1}$
that has a domain with a cardinality polynomial in $n$,
and this follows from \cite[Section 1.2]{LubyW05}.
The idea is to represent the distribution by a randomization over a family of 2-universal hash functions that
assigns each bidder u.a.r.~values from the set $\{t_1,\dots,t_{2\Delta}\}$ and to consider the subset of bidders with value $t_1$.
Furthermore, this implies that also a distribution over the random sets $B^c$ computed by Algorithm~\ref{alg:randset} exists
that has a domain with a cardinality polynomial in~$n$.

\begin{lemma}[\cite{LubyW05}]
Given a set $B$ with $|B|=n$, for any integer $1\le \Delta \le n$,
there exists a pairwise independent distribution over subsets of $B$ with probability $1/2^{\lceil \log_2 \Delta\rceil +1}$
and a domain with a cardinality in $\mathcal{O}(n^2)$.
\end{lemma}

Thus, instead of picking a subset from a pairwise independent distribution with probability $1/2^{\lceil \log_2 \Delta\rceil +1}$
in Algorithm~\ref{alg:randset},
we can iterate over the domain of the distribution in polynomial time.
This gives us at least the same social welfare.
Moreover, note that 
$1/(2\Delta)\geq 1/2^{\lceil \log_2 \Delta\rceil +1} > 1/(4\Delta)$.
By slightly modifying the proof of Lemma \ref{lem:randset}, we can show that
every bidder is in $B^c$ with a probability of at least $3/(16\Delta)$.

\begin{corollary}\label{coro:deterministic}
A deterministic $\alpha$-approximation algorithm $f$ for combinatorial auctions without conflicts
can be turned into
a deterministic
$(16\Delta\alpha/3)$-approx.\ algorithm $f^c$ for combinatorial auctions with $\Delta$ bounded and $G_I$ arbitrary.
\label{cor:detalg}
\end{corollary}

Furthermore, we can extend the results to \emph{maximal-in-range} algorithms (see \cite{DobzinskiD13}) which are important for the design of truthful approximation mechanisms. In fact, most deterministic truthful approximation mechanisms for combinatorial auctions are maximal-in-range mechanism. 

\begin{definition}
An algorithm $f$ is called ``maximal-in-range'' if there exists a subset $A'$ of the set of allocations $A$ for which $f(v_1,\dots,v_n)\in \arg\max_{S\in A'}(\sum_{i\in B} v_i(S))$.
\end{definition}

Given a maximal-in-range algorithm $f$ for CA without conflicts that is a deterministic $\alpha$-approximation algorithm,
we show in Algorithm~\ref{alg:affine} how to construct a maximal-in-range algorithm $f^c$ for CA with conflicts
that is a deterministic $(16\Delta\alpha/3)$-approximation algorithm.
Hence, Theorem~\ref{thm:randmain1} follows. Note that Algorithm~\ref{alg:affine} calls $f$ always for the same set of bidders and set of items,
only the valuations of the bidders change; thus, the target set $A'$ of $f$ is the same in each call.

\begin{algorithm}[h]
\SetAlgoLined
Let $D$ be the domain of a distribution over the random set $B^c$ computed by Algorithm~\ref{alg:randset} satisfying $|D|\in\mathcal{O}(n^2)$\;
Let $A'\subseteq A$ be the target set of $f$\;
Set $\OPT\gets (\emptyset,\dots,\emptyset)$\;
\ForEach{$B'\in D$}
{
\lForAll{$i\in B$}{set $v^{B'}_i\gets v_i$ if $i\in B'$ and $v^{B'}_i\gets 0$ else\;}
Set $\OPT(B')\gets f(v^{B'})$\;
\If{$\sum_{i\in B}v^{B'}_i(\OPT_i(B'))\ge\sum_{i\in B} v_i(\OPT_i)$}
{Set $\OPT$ to the following assignment:\newline (1)~all $i\in B'$ get the same items as in $\OPT(B')$; and (2)~others get no items\;
}
}
\KwRet{$\OPT$}
\caption{Maximal-in-range algorithm $f^c$ over range $A'_D$.}
\label{alg:affine}
\end{algorithm}

\begin{pfof}{Theorem \ref{thm:randmain1}}

Let us assume that $\OPT$ was set to its final value when ${B'}=B^*$. Furthermore, assume that $B^c$ is a random subset computed by Algorithm~\ref{alg:randset}. It follows that
\begin{multline*}
\sum_{i\in B} v_i(\OPT_i)=\sum_{i\in B} v^{B^*}_i(\OPT_i)=\sum_{i\in B} v^{B^*}_i(\OPT_i({B^*}))\\ \ge \mathbf{E}_{B^c} \bigg[\sum_{i\in B} v^{B^c}_i(\OPT_i({B^c}))\bigg] \overset{(*)}{\ge} \mathbf{E}_{B^c} \bigg[\sum_{i\in B} v^{B^c}_i(\OPT_i(B))\bigg]\\ =\sum_{i\in B} v_i(\OPT_i(B)) \cdot\Pr{i\in {B^c}}\ge \frac{3}{16\Delta} \cdot \sum_{i\in B} v_i(\OPT_i(B))\,.
\end{multline*}
Inequality $(*)$ holds because $\OPT({B^c})\in \arg\max_{S\in A'}\sum_{i\in B} v^{B^c}_i(S_i)$ and $\OPT(B)\in A'$.
Since the maximum social welfare is at most $\alpha\cdot \sum_{i\in B} v_i(\OPT_i(B))$ the claimed approximation factor follows.

We still have to show that the algorithm is maximal-in-range. 
For each $S\in A'$ and ${B'}\in D$ let $S^{B'}$ be the assignment when all bidders in $i\in {B'}$ get the same set of items as in $S$ and all other bidders get no items. Define $A'_D:=\{S^{B'}|(S,{B'})\in A'\times D\}$. We show next that $f^c$ is maximal on the subset $A'_D\subseteq A$. It holds that
$$
\sum_{i\in B}v_i(\OPT_i)=\max_{{B'}\in D}\sum_{i\in B}v^{{B'}}_i(\OPT_i({B'})) =\max_{{B'}\in D}\max_{S\in A'}\sum_{i\in B}v_i^{{B'}}(S_i)=\max_{S^{B'}\in A'_D}\sum_{i\in B}v_i(S_i^{{B'}})\,.
$$
The pricing scheme for a truthful mechanism is given in \cite[Proposition 9.31]{Nisan07}.
\end{pfof}


\subsection{Bounded Out-degree in the Item Conflict Graph}\label{subsect:sampling-item}


Here we provide the proof of Theorem~\ref{thm:randmain2}, which follows from Theorem~\ref{thm:min-of-two-Delta} below.
Recall that we here handle the case when $\Delta_I$ is bounded, $G$ is arbitrary, and valuations are fractionally sub-additive.
Again, the results apply to arbitrary restrictions on the valuations (e.g., submodular valuations).

\begin{theorem}\label{thm:min-of-two-Delta}
Suppose that valuations are fractionally sub-additive, and that the original approximation guarantees of $f^c$ in Theorem~\ref{thm:detalg}, Corollary~\ref{coro:truthful}, Corollary~\ref{cor:detalg}, and Theorem~\ref{thm:randmain1} are in the form of $C\Delta\alpha$,
where $C$ is either $4$ or $16/3$.
Then $f^c$ can be modified so that the approximation guarantees are changed to $C\Delta_I\alpha$.
\end{theorem}



\begin{pfof}{Theorem~\ref{thm:min-of-two-Delta} (for Lemma~\ref{thm:detalg} with bounded $\Delta_I$)}
The idea is to restrict the item set $I$ to a random set $I^c$ that is independent of the valuations of the bidders and then to call the $\alpha$-approximation algorithm for the restricted item set $I^c$.
Note that we can use Algorithm~\ref{alg:randset} also for items. Thus, by the same arguments as for bidders we obtain a random set of items $I^c$ where items have no conflicts and where every item is in $I^c$ with probability at least $1/(4\Delta_I)$. Let $(S_1(I'),\dots,S_n(I'))$ be the allocation that $f$ returns when we restrict the set of items to $I'\subseteq I$. Furthermore, let $(\OPT_1(I'),\dots,\OPT_n(I'))$ be the optimal allocation of the item set $I'$. It holds that
\begin{multline*}
\mathbf{E}_{I^c}\bigg[\sum_{i\in B}v_i(S_i(I^c))\bigg]
\ge \mathbf{E}_{I^c}\bigg[\frac{1}{\alpha}\sum_{i\in B}v_i(\OPT_i(I^c))\bigg]
\\ \overset{(*)}{\ge} \mathbf{E}_{I^c}\bigg[\frac{1}{\alpha}\sum_{i\in B}v_i(\OPT_i(I)\cap I^c)\bigg]
\ge \frac{1}{\alpha}\sum_{i\in B}\mathbf{E}_{I^c}\left[v_i(\OPT_i(I)\cap I^c)\right]
\\ \overset{(**)}{\ge} \frac{1}{4\Delta_I\alpha}\sum_{i\in B}\mathbf{E}_{I^c}\left[v_i(\OPT_i(I))\right]\,.
\end{multline*}
Above, inequality $(*)$ follows because $\OPT_i(I^c)$ is optimal for item set $I^c$, and inequality $(**)$ follows by Proposition~\ref{prop:feige}.
\end{pfof}

\begin{pfof}{Theorem~\ref{thm:min-of-two-Delta} (for Corollary~\ref{cor:randmech} with bounded $\Delta_I$)}
Again, we restrict the item set $I$ to a random set $I^c$ as in Theorem~\ref{thm:detalg} and then to call mechanism $(f,p)$ for the restricted item set. The approximation guarantee follows from Theorem~\ref{thm:detalg} and universally truthfulness (resp.\ truthfulness in expectation) follows since truthful bidding is a dominant strategy for each realization of $I^c$.
\end{pfof}

\begin{pfof}{Theorem~\ref{thm:min-of-two-Delta} (for Lemma~\ref{cor:detalg} with bounded $\Delta_I$)}
Note that in Theorem~\ref{thm:detalg} we use the same randomization technique for $B^c$ and for $I^c$.
Thus, we can apply our derandomization technique for $B^c$ also to $I^c$.
\end{pfof}

\begin{algorithm}[h]
\SetAlgoLined
Let $D$ be the domain of a distribution over the random set $I^c$ computed by Algorithm~\ref{alg:randset} satisfying $|D|\in\mathcal{O}(m^2)$\;
Let $A'\subseteq A$ be the target set of $f$\;
Set $\OPT\gets (\emptyset,\dots,\emptyset)$\;
\ForEach{$I'\in D$}
{
Define $v^{I'}_i(S):=v_i(S\cap I')$ for all $S\subseteq I$ and $i\in B$\;
Set $\OPT(I')\gets f(v^{I'})$\;
\lIf{$\sum_{i\in B}v^{I'}_i(\OPT_i(I'))\ge\sum_{i\in B} v_i(\OPT_i)$}
{$\forall i\in B:\ \OPT_i\gets \OPT_i(I')\cap I'$\;
}
}
\KwRet{$\OPT$}
\caption{Maximal-in-range algorithm $f^c$ over range $A'_D$ when $\Delta_I$ is bounded.}
\label{alg:itemaffine}
\end{algorithm}

\begin{pfof}{Theorem~\ref{thm:min-of-two-Delta} (for Theorem~\ref{thm:randmain1} with bounded $\Delta_I$)}
Given a maximal-in-range algorithm $f$ for combinatorial auctions without conflicts that is a deterministic $\alpha$-approximation algorithm we show in Algorithm~\ref{alg:itemaffine} how to construct a maximal-in-range algorithm $f^c$ for combinatorial auctions with conflicts that is a deterministic $(16\Delta_I\alpha/3)$-approximation algorithm. As in Algorithm~\ref{alg:affine}, Algorithm~\ref{alg:itemaffine} always calls $f$ for the same set of bidders and set of items; thus, the target set $A'$ of $f$ is the same in each call.

Let us assume that $\OPT$ was set to its final value when ${I'}=I^*$. Furthermore, assume that $I^c$ is a random subset computed by Algorithm~\ref{alg:randset}. It follows that
\begin{multline*}
\sum_{i\in B}v_i(\OPT_i)
=\sum_{i\in B}v^{I^*}_i(\OPT_i(I^*))
\ge \mathbf{E}_{I^c}\bigg[\sum_{i\in B}v^{I^c}_i(\OPT_i(I^c))\bigg]\\
\overset{(*)}{\ge} \mathbf{E}_{I^c}\bigg[\sum_{i\in B}v^{I^c}_i(\OPT_i(I))\bigg]
= \sum_{i\in B}\mathbf{E}_{I^c}\left[v_i(\OPT_i(I)\cap I^c)\right]
\\ \overset{(**)}{\ge} \frac{3}{16\Delta_I}\sum_{i\in B}v_i(\OPT_i(I))\,.
\end{multline*}
Inequality $(*)$ holds because $\OPT(I^c)\in \arg\max_{S\in A'}\sum_{i\in B} v^{I^c}_i(S_i)$ and $\OPT(I)\in A'$, and Inequality $(**)$ follows from Lemma~\ref{prop:feige}. Since the maximum social welfare it at most $\alpha\cdot \sum_{i\in B} v_i(\OPT_i(I))$, the claimed approximation guarantee follows.

We still have to show that the algorithm is maximal-in-range. 
For each $S\in A'$ and ${I'}\in D$ let $S^{I'}:=(S_1\cap I',\dots, S_n\cap I')$. Define $A'_D:=\{S^{I'}|(S,{I'})\in A'\times D\}$. We show that $f^c$ is maximal on the subset $A'_D\subseteq A$. It holds that
$$
\sum_{i\in B} v_i(\OPT_i)
=\max_{I'\in D}\sum_{i\in B} v^{I'}_i(\OPT_i(I'))
=\max_{I'\in D}\max_{S\in A'}\sum_{i\in B} v^{I'}_i(S_i)
=\max_{S^{I'}\in A'_D}\sum_{i\in B} v_i(S^{I'}_i)\,.
$$
Again, the pricing scheme for a truthful mechanism is given in \cite[Proposition 9.31]{Nisan07}.
\end{pfof}


%% file: cone_program_appendix.tex
\section{Proofs and Results Omitted in Section \ref{sect:CP}}\label{sect:CP-appendix}

\subsection{Solving Dual of \textsf{CPR-C} in Poly-time using Ellipsoid Algorithm}\label{subsect:CP-solve-in-ptime}

First, we refer the readers to \cite[Chapters 2--4]{GroetschelLovaszSchrijver1988} for details of ellipsoid algorithm.
We will use the following result from fundamental linear algebra:

\begin{lemma}[{\cite[Section 0.1]{GroetschelLovaszSchrijver1988}}]\label{lem:spsd-property}
Let $M$ be a symmetric $q\times q$ real matrix.
$M$ is positive semi-definite if and only if
there exists $w_1,w_2,\cdots,w_q\in \mathbb{R}^q$ such that for $1\leq k,\ell \leq q$, $M_{k\ell} = w_k \cdot w_\ell$.
Furthermore, $M$ is positive definite if and only if the vectors $w_1,w_2,\cdots,w_q$ are linearly independent.
\end{lemma}

For our problem, we define the following proper cone $K$:
$K$ consists of all points $\left(\{\xiS\},\{\alpha_i\},\{\beta_k\},M\right)$,
where $\{\xiS\},\{\alpha_i\},\{\beta_k\}$ are vectors of dimension $(2^m-1)n,~n,~m$ respectively,
and all of them have non-negative entries;
$M$ is symmetric positive semi-definite $(n+1)\times (n+1)$ real matrix.
It is easy to verify that $K$ is a proper cone. By following the arguments in \cite[Examples 2.23 and 2.24]{BoydV04},
it is easy to show that $\Ks = K$, i.e., $K$ is self-dual.

Using Lemma \ref{lem:spsd-property} and introducing slack variables $\alpha$'s and $\beta$'s,
we can rewrite \textsf{CPR-C} in the standard CP form \textsf{CP-STD}:
\begin{align*}\tag{\textsf{CPR-C'}}
\min ~ & ~-\sum_{i\in B} \sum_{S\neq\emptyset} v_i(S) \cdot \xiS & & & & \\
\mbox{s.t.}~ & -\sum_{S\neq\emptyset} \xiS - \alpha_i &=& -1, & \forall i\in B & ~~~~~\mathbf{(u_i)}\\
& -\sum_{S\ni k} \sum_{i\in B} \xiS - \beta_k &=& -1, & \forall k\in I & ~~~~~\mathbf{(p_k)}\\
& 2\sum_{S\neq\emptyset} \xiS - M_{0i} &=& ~1, & \forall i\in B & ~~~~~\mathbf{(z_i)}\\
& M_{0i} + M_{0j} + M_{\min\{i,j\}~\max\{i,j\}} &=& -1, & \forall (i,j)\in E & ~~~~~\mathbf{(y_{ij})}\\
& M_{00} &=& ~1, & & ~~~~~\mathbf{(q_0)}\\
& M_{ii} &=& ~1, & \forall i\in B & ~~~~~\mathbf{(q_i)}\\
& \left(\{\xiS\},\{\alpha_i\},\{\beta_k\},M\right) & \succeq_K & ~0.
\end{align*}

Instead of solving \textsf{CPR-C'} directly, we will solve its dual.
Each equality constraint in \textsf{CPR-C'} will associate to a variable in the dual.
We have written the variables down on the right of their corresponding constraints.

To ensure that solving the dual of \textsf{CPR-C'} is equivalent to solving \textsf{CPR-C'},
we need to check that strong duality holds between them. We do this soon later in this appendix.

The dual of \textsf{CPR-C'} is
\begin{align*}\tag{\textsf{CPR-C-DUAL}}
\max ~ & ~ -\sum_{i\in B} u_i - \sum_{k\in I} p_k + \sum_{i\in B} z_i - \sum_{(i,j)\in E} y_{ij} + q_0 + \sum_{i\in B} q_i & \\
\mbox{s.t.}~ & ~ v_i(S) - u_i - \sum_{k\in S} p_k + 2 z_i \leq 0, & \forall i\in B,S\subseteq I\\
& ~ u_i \geq 0, & \forall i\in B\\
& ~ p_k \geq 0, & \forall k\in I\\
& ~ -Q~\mbox{is SPSD},
\end{align*}
where $Q$ is the symmetric $(n+1)\times (n+1)$-matrix determined as follows:
\begin{align*}
Q_{ii} &= ~q_i, & & \forall i\in B\cup\{0\} \\
Q_{0i} = Q_{i0} &= ~-z_i + \sum_{j:(i,j)\in E} y_{ij} + \sum_{j:(j,i)\in E} y_{ji}, & & \forall i\in B\\
\\
Q_{ij} = Q_{ji} &= ~\begin{cases}
0 & ~\mbox{if }(i,j)\notin E~\mbox{and}~(j,i)\notin E\\
y_{ij} &~\mbox{if }(i,j)\in E~\mbox{and}~(j,i)\notin E\\
y_{ji} &~\mbox{if }(i,j)\notin E~\mbox{and}~(j,i)\in E\\
y_{ij} + y_{ji} &~\mbox{if }(i,j)\in E~\mbox{and}~(j,i)\in E
\end{cases},
& & \forall~\mbox{distinct}~i,j\in B.
\end{align*}

The final step is to design a poly-time separation oracle:
\begin{itemize}
\item If $u_i<0$ for some $i\in B$ or $p_k<0$ for some $k\in I$, we have an obvious separation hyperplane.
\item Since the dimension of $-Q$ is \textsf{poly}$(n)$,
we can use a standard algorithm to check whether it is SPSD in \textsf{poly}$(n)$ time,
and obtain a separation hyperplane if $-Q$ is not SPSD. See \cite[Example 2]{OliveiraM14} for details.
\item If $v_i(S) - u_i - \sum_{k\in S} p_k + 2z_i > 0$ for some $i\in B$ and $S\subseteq I$,
then we can use the demand oracle of bidder $i$ to find $S=S^*$ that maximizes $v_i(S) - \sum_{k\in S} p_k$.
Then $v_i(S^*) - u_i - \sum_{k\in S^*} p_k + 2z_i > 0$, which provides us a separation hyperplane.
This is almost identical to the separation oracle used in the ellipsoid algorithm for solving \textsf{LPR-NC} in \cite{NisanS06}.
\end{itemize}

\subsection{Strong Duality of \textsf{CPR-C}}\label{subsect:CP-strong-duality}

While strong duality always holds for LP, it does not always hold for CP.
To check strong duality, we verify that the primal program \textsf{CPR-C'} satisfies \emph{Slater's condition},
i.e., find a feasible point which satisfies all equality constraints, and \emph{strictly satisfy} the conic constraint.
In other words, we need to find a feasible point $\left(\{\xiS\},\{\alpha_i\},\{\beta_k\},M\right)$
which satisfies all equality constraints, and such that $\xiS > 0$ for all $i\in B,S\subseteq I$,
$\alpha_i > 0$ for all $i\in B$, $\beta_k > 0$ for all $k\in I$, and $M$ is positive \emph{definite}.

Here, we only consider the cases $n\geq 2$; the auction with $n=1$ bidder is trivial.

Consider the point with $\xiS = \frac{1}{4(2^m-1)n^2}$ for all $i\in B$ and $S\subseteq I$.
Then $\alpha_i = 1 - (2^m-1)\cdot \frac{1}{4(2^m-1)n^2} > 0$ and
$\beta_k = 1 - 2^{m-1}n\cdot \frac{1}{4(2^m-1)n^2} > 0$.
Also, $\forall i\in B$, $M_{0i} = 2(2^m-1) \cdot \frac{1}{4(2^m-1)n^2} - 1 = \frac{1}{2n^2}-1$,
and $\forall i,j\in B$ where $i\neq j$, we choose $M_{ij} = -1 - M_{0i} - M_{0j} = 1 - \frac{1}{n^2}$.
Recall that $\forall i\in B$, $M_{00} = M_{ii} = 1$.

For notational convenience, let $\ep = \frac{1}{2n^2}$, i.e., $\forall i\in B$, $M_{0i} = \ep - 1$;
$\forall i,j\in B$ where $i\neq j$, $M_{ij} = 1-2\ep$.
By Lemma \ref{lem:spsd-property}, to check that $M$ is positive definite, equivalently,
we find linearly independent $w_0,w_1,\cdots,w_n \in \mathbb{R}^{n+1}$
such that $\forall i,j\in B\cup\{0\}$, $w_i \cdot w_j = M_{ij}$.

Let
\begin{align*}
w_0 &= (1,0,0,\cdots,0)\\
w_1 &= (\ep-1,a_1,0,0,\cdots,0)\\
w_2 &= (\ep-1,b_1,a_2,0,0,\cdots,0)\\
w_3 &= (\ep-1,b_1,b_2,a_3,0,0,\cdots,0)\\
& \vdots \\
w_n &= (\ep-1,b_1,b_2,b_3,\cdots,b_{n-1},a_n).
\end{align*}
Note that $\forall i\in B\cup\{0\}$, $w_i$ has $n-i$ trailing zeroes.
Also, $\forall i\in B$, the first entry of $w_i$ is $\ep-1$, followed by $b_1,b_2,\cdots,b_{i-1}$,
and then followed by $a_i$ and the trailing zeroes.
These ensure that $w_0 \cdot w_0 = 1 = M_{00}$ and $w_0\cdot w_i = \ep-1 = M_{0i}$ for all $i\in B$.

We will determine $a_1,b_1,a_2,b_2,a_3,b_3,\cdots,a_{n-1},b_{n-1},a_n$ in this order.
We will show that for every $i\in B$, $a_i \geq \frac{\sqrt{133}}{12n}$,
and $b_i$'s are negative with $|b_i| \leq \frac{1}{3n^3}$.

Since $w_1\cdot w_1 = 1$, $(\ep-1)^2 + (a_1)^2 = 1$ and thus
$$a_1 = \sqrt{2\ep-\ep^2} = \sqrt{\frac{1}{n^2} - \frac{1}{4n^4}} = \sqrt{\frac{1}{n^2}\left(1-\frac{1}{4n^2}\right)} \geq \sqrt{\frac{15}{16 n^2}} > \frac{\sqrt{133}}{12n}.$$
Since for $i>1$, $w_1\cdot w_i = 1-2\ep$, $(\ep-1)^2+a_1 b_1 = 1-2\ep$. Hence $a_1 b_1 = -\ep^2$
and thus $b_1$ is negative with
$$|b_1| = \ep^2 / a_1 = \frac{1}{4n^4 a_1} < \frac{3}{\sqrt{133}n^3} < \frac{1}{3n^3}.$$

Next, we proceed by induction.
Suppose that for some $q\geq 1$, $a_1,a_2,\cdots,a_q \geq \frac{\sqrt{133}}{12n}$
and $b_1,b_2\cdots,b_q$ are negative with $|b_1|,|b_2|\cdots,|b_q| \leq \frac{1}{3n^3}$.

Since $w_{q+1} \cdot w_{q+1} = 1$, $(\ep-1)^2 + \sum_{\ell=1}^q (b_\ell)^2 + (a_{q+1})^2 = 1$ and thus
\begin{align*}
a_{q+1} = \sqrt{2\ep - \ep^2 - \sum_{\ell=1}^q (b_\ell)^2} \geq \sqrt{\frac{1}{n^2} - \frac{1}{4n^4} - n\cdot \frac{1}{9n^6}}\\
~~~~~\geq \sqrt{\frac{1}{n^2} - \frac{1}{4n^4} - \frac{1}{18n^4}} = \sqrt{\frac{1}{n^2}\left(1-\frac{11}{36n^2}\right)} \geq \frac{\sqrt{133}}{12n}.
\end{align*}
Since for $i>q+1$, $w_{q+1}\cdot w_i = 1-2\ep$, $(\ep-1)^2 + \sum_{\ell=1}^q (b_\ell)^2 + a_{q+1} b_{q+1} = 1-2\ep$.
Hence
$$0 > a_{q+1} b_{q+1} = -\ep^2 - \sum_{\ell=1}^q (b_\ell)^2 \geq -\frac{1}{4n^4} -n\cdot \frac{1}{9n^6} \geq -\frac{1}{4n^4} - \frac{1}{18n^4} = -\frac{11}{36n^4},$$
and thus $b_{q+1}$ is negative with
$$|b_{q+1}| \leq \frac{11}{36n^4 a_{q+1}} \leq \frac{11}{36n^4} \cdot \frac{12n}{\sqrt{133}} = \frac{11}{3\sqrt{133}n^3} < \frac{1}{3n^3}.$$
These complete the induction.

Since all $a_i$'s are strictly positive, $w_0,w_1,w_2,\cdots,w_n$ are linearly independent.